\documentclass[floatfix,aps,twocolumn,preprintnumbers,amsmath,amssymb,nofootinbib,superscriptaddress]{revtex4}

\usepackage{graphicx,color}
\usepackage{amsmath,amssymb,bm}
\usepackage{dcolumn}
\usepackage{comment}

\newcommand{\bea}{\begin{eqnarray}}
\newcommand{\eea}{\end{eqnarray}}
\newcommand{\be}{\begin{equation}}
\newcommand{\ee}{\end{equation}}
\newcommand{\np}{{\bf p}}

\newcommand{\nr}{{\bf r}}

\newcommand{\nh}{{\bf h}}

\newcommand{\nk}{{\bf k}}
\newcommand{\nl}{{\bf l}}

\newcommand{\nq}{{\bf q}}

\newcommand{\nj}{{\bf j}}

\newcommand{\nK}{{\bf K}}
\newcommand{\nA}{{\bf A}}
\newcommand{\nB}{{\bf B}}
\newcommand{\na}{{\bf a}}
\newcommand{\nb}{{\bf b}}
\newcommand{\Qbar}{\not{\!Q}}

\newcommand{\kbar}{\not{\!k}}
\newcommand{\Pbar}{\not{\!P}}
\newcommand{\pbar}{\not{\!p}}

\newcommand{\nsigma}{\mbox{\boldmath $\sigma$}}
\newcommand{\ntau}{\mbox{\boldmath $\tau$}}

\def\XXint#1#2#3{{\setbox0=\hbox{$#1{#2#3}{\int}$}
     \vcenter{\hbox{$#2#3$}}\kern-.5\wd0}}

\def\1{\'{\i}}

\begin{document}

\title{Interference between Meson Exchange and One-Body Currents 
 in Quasielastic Electron Scattering }

\author{P.R. Casale} \email{palomacasale@ugr.es}
\affiliation{Departamento de
  F\'{\i}sica At\'omica, Molecular y Nuclear}
\affiliation{Instituto Carlos I
  de F{\'\i}sica Te\'orica y Computacional Universidad de Granada,
  E-18071 Granada, Spain.}

\author{J.E. Amaro}\email{amaro@ugr.es} 
\affiliation{Departamento de
  F\'{\i}sica At\'omica, Molecular y Nuclear}
\affiliation{Instituto Carlos I
  de F{\'\i}sica Te\'orica y Computacional Universidad de Granada,
  E-18071 Granada, Spain.}

\author{V. Belocchi}
\affiliation{Dipartimento di Fisica Universit\`a di Torino, P. Giuria 1, 
10125 Torino, Italy}
\affiliation{INFN Sezione di Torino, 10125 Torino, Italy}
\affiliation{Instituto de F\'isica Corpuscular (IFIC), Consejo Superior de Investigaciones
Cient\'ificas (CSIC) and Universidad de Valencia, E-46980 Paterna, Valencia, Spain}

\author{M.B. Barbaro}
\affiliation{Dipartimento di Fisica Universit\`a di Torino, P. Giuria 1, 
10125 Torino, Italy}
\affiliation{INFN Sezione di Torino, 10125 Torino, Italy}

\author{A De Pace}
\affiliation{INFN Sezione di Torino, 10125 Torino, Italy}
  
\author{M. Martini}
  \affiliation{
IPSA-DRII, 63 boulevard de Brandebourg, 94200 Ivry-sur-Seine, France}
\affiliation{
Sorbonne Universit\'e, CNRS/IN2P3,\\
Laboratoire de Physique Nucl\'eaire et de Hautes Energies (LPNHE), 75005 Paris, France
}

\date{\today}

\begin{abstract}

In this work, we present a detailed analysis of the interference
between meson exchange currents (MEC) and one-body currents in
quasielastic electron scattering, with a focus on the sign of this
interference in the transverse response for one-particle emission. We
prove that the interference of both the Delta and pion-in-flight
currents with the one-body current is negative, leading to a partial
cancellation with the seagull current. This is mathematically
demonstrated within the framework of the Fermi gas model. By comparing
these interferences across various independent particle models, both
relativistic and non-relativistic, our results indicate that all
studied models display the same behavior.

\end{abstract}


\keywords{Electron scattering, Meson-exchange currents, $\Delta$ resonance, Fermi gas, Shell model. Relativistic Mean Field}

\maketitle

\section{Introduction}

Quasielastic electron  scattering plays a pivotal role in probing
nuclear structure and dynamics \cite{Cen01,Bof96}. This scattering
process, wherein an electron scatters off a nucleon causing it to be
ejected from the nucleus, provides a wealth of information about the
underlying nuclear response functions. These response functions
provide insight into the distribution of charges and currents in the
nucleus and the dynamics of nucleons within the nuclear medium
\cite{Mon71,Whi74,Ros80,Hor90,Weh93,Sar93,Jou96,Car02}.

The analysis of current accelerator-based neutrino experiments
necessitates a thorough understanding of the probability of neutrino
interactions with nuclei
\cite{Alv14,Mos16,Kat17,Ank17,Ben17,Alv18,Ama20,Ank22,Alv25}.  Given the
challenges in obtaining precise measurements of neutrino-nucleus
cross-sections, nuclear models of reactions, such as the
$(\nu_{\mu},\mu^-)$ charge-changing processes, are
indispensable. Neutrino-induced reactions are closely related to
electron-induced reactions; in both cases, the electroweak current is
explored ---within the weak sector for neutrinos and the
electromagnetic sector for electrons. Therefore, the same nuclear
models used to describe electron scattering can, in principle, be
extended to the case of neutrino scattering by modifying the nuclear
current operator. A significant contribution to $(\nu_{\mu},\mu^-)$
process arises from the quasielastic region, dominated by
single-particle emission, although the importance of two-particle
emission has also been recognized. In this study, we focus on the
single-particle emission, excluding pion
production and inelastic processes.  

Analyses of electron and neutrino scattering data have indicated the
necessity of including mechanisms that enhance the transverse response
\cite{Bos12,Bod22}, with meson-exchange currents (MEC) identified as a
potential source \cite{Car02,Lov16}. MEC, being two-body operators,
can induce both single-particle (1p1h) and two-particle (2p2h)
excitations within independent particle models (Fermi gas, Mean field). 
The specific 2p2h channel has been extensively studied
\cite{Van80,Alb84,Dek94,Pac03,Mar09,Mar11,Nie11,Meg15,Meg16,Van17},
demonstrating an important enhancement of the transverse response in
the dip region (between the QE and the $\Delta$ peaks). While
discrepancies exist among various approaches, there is a consensus
that 2p2h emission can enhance the inclusive CCQE neutrino cross
section about 15 to 20\%, thus achieving better agreement with
experimental data when these effects are considered alongside pion
emission.

The effect of MEC in the 1p1h channel, where one-body (1b) currents
interfere with two-body (2b) currents \cite{Ama98,Ama02,Ama02b}, 
has received less attention in
the modeling of neutrino scattering and is more controversial.
Independent particle models (IPM) ---Fermi gas, mean field--- typically predict
a small and negative MEC effect due to a cancellation 
between the positive seagull  and negative pion-in-flight (or pionic)
and $\Delta$ currents in the transverse response. At intermediate
momentum transfers ($q \sim 500 MeV/c$), the $\Delta$ current dominates,
leading to a small reduction in the transverse response. This
reduction arises because, in the matrix element of the MEC between the
ground state and a 1p1h excitation, the direct term is negligible or
zero, with the exchange term causing a net reduction
\cite{Koh81,Alb90,Ama94a,Ama94b,Ama03,Cas23}.

In contrast, calculations by Fabrocini \cite{Fab97} in nuclear matter
within the correlated basis function (CBF) theory have shown a
positive effect on the transverse response. This positive effect was
identified as due to tensor correlations between nucleons
\cite{Lei90,Car02}.  When these correlations are omitted in
Fabrocini's calculations, the results align with those of independent
particle models.

It should be noted that Fabrocini's calculation includes only 1p1h
excitations. There are, however, other calculations, such as the
Green's Function Monte Carlo (GFMC) calculations by Carlson et
al. \cite{Car02,Lov14}, which provide an accurate description of the
wave function by incorporating short-range correlations and show an
enhancement of the transverse response due to MEC. Interestingly, in
this model, the enhancement persists even when the ground-state wave
function includes only central correlations (i.e., without tensor
correlations). According to Ref. \cite{Car02}, this is a consequence
of the complete set of final states included in the Euclidean response
functions. Therefore, within the GFMC approach, 
all final states ---in particular both 1p1h and 2p2h--- are included, and
it is not possible to
isolate the contribution from specific final states such as 1p1h.

One of the motivations for the present work is that some recent
calculations  using independent-particle \cite{Fra23,Fra25} or  spectral function \cite{Lov23} models have reported
 a large and positive interference of one-body and
two-body currents in the transverse response, without the need for
tensor correlations in the nuclear wave function. These results
contradict previous calculations mentioned earlier, and their origin
is not entirely understood, necessitating further clarification. A
detailed study, such as the one presented in this article, is required
to systematically analyze the theoretical foundations of the
interference response and to assess whether the observed discrepancies
arise from fundamental differences in the modeling of nuclear dynamics
or from specific approximations used in these calculations.

To that end in this article we demonstrate in detail two results for low
momentum transfer.  The low momentum results will help shed light on
this matter, as models based on similar assumptions should
approximately coincide under non-relativistic conditions for low
momentum transfer.  Specifically, we will show that the interference
of the $\Delta$ current with the one-body current in the transverse
response is negative within the non-relativistic Fermi gas model.
Similarly, the interference of the pion-in-flight current with the
one-body current is also negative, leading to a partial cancellation
with the positive interference of the seagull current.  We will
demonstrate the results in detail, including all relevant formulas
and, where applicable, analytical results, ensuring that our
calculations with the Fermi gas will be reproducible. This is
scientifically desirable to avoid any reasonable doubt.

We will then show, through calculations using a series of independent
particle models (IPM), including relativistic Fermi gas (RFG) mean
field, plane wave approximation, spectral function, relativistic mean
field, and Dirac-equation based shell model, that all of them yield
similar results for the 1b2b interference transverse response.  This
demonstrates that all these models without tensor correlations do not
violate the results for low to intermediate values of the momentum
transfer. 

We proceed systematically in Section II introducing the formalism of
electron scattering, starting with the relativistic MEC model of
ref. \cite{Cas23}.  By taking the non-relativistic limit, we first obtain
the MEC expressions for low energy-momentum, 
verifying that we arrive at the
non-relativistic Riska's expressions \cite{Ris89}, which are the
standard operators typically used in calculations including the seagull,
pion-in-flight, and Delta currents \cite{Fru84,Eri88,Ris79}.

At the end of Section II we demonstrate two low momentum results that establish the 
negative interference of the $\Delta$ and pion-in-flight currents with the one-body current. We compute in
detail the 1p1h matrix element of the non-relativistic MEC between
plane waves, providing analytical expressions 
after performing explicitly the spin sums.
Furthermore, we will derive the formulas for the one-body--two-body (1b2b) interferences,
explicitly demonstrating that the pion-in-flight and Delta
contributions are negative.

In the results Section III we will compare the interference responses
calculated with various independent particle models
(IPM). Specifically, we will consider the relativistic and
non-relativistic Fermi gas models, the relativistic mean field of
nuclear matter, the mean field with Woods-Saxon potential, the Dirac
equation-based potential, and the plane wave approximation
(PWA),
illustrating the similarities and differences in the OB-MEC transverse
interference response. 
 Additionally, we will examine the superscaling model and the
spectral function model. Finally, in Section IV, we will present our
conclusions.

\section{Formalism}

In this section, we present the formalism of electron scattering and
the current operators, including both one-body and meson exchange
currents (MECs). We begin with the relativistic expressions and
proceed to the non-relativistic limit, which will be applied in the
low momentum transfer kinematics. We maintain a detailed level of
discussion, providing many mathematical details that are already
known, with the aim of making the content understandable to a high
proportion of interested readers who may not be experts in the field.
This didactic component is intended to ensure that our results are
reproducible by anyone who wishes to do so.

\subsection{Response functions}

The starting point is the inclusive electron scattering cross section
in plane-wave Born approximation with one photon exchange. The
exchanged photon is virtual, carrying an energy transfer \(\omega\)
and a momentum transfer \(\nq\) to the nucleus. We choose the $z$-axis
in the direction of $\nq$. The initial electron energy is
\(\epsilon\), and the final electron exits the interaction region with
a scattering angle \(\theta\) and an energy \(\epsilon' = \epsilon -
\omega\). As usual, we use units where $\hbar=c=1$.  The
double-differential cross section
can be written as
\begin{equation}  \label{cross}
\frac{d\sigma}{d\Omega d\epsilon'}
= \sigma_{\rm M}
\left[
\frac{Q^{4}}{q^{4}} R_L + 
\left(\tan^2\frac{\theta}{2}-\frac{Q^{2}}{2q^{2}}\right)
 R_T \right],
\end{equation}
where $\sigma_{\rm M}$ is the Mott cross section and 
$Q^{2}=\omega^{2}-q^{2}<0$ is the square of the four-momentum transfer.  
The longitudinal and transverse response functions, \(R_L(q, \omega)\)
and \(R_T(q, \omega)\), depend only on $q=|\nq|$ and $\omega$, and
are defined as the following components of the hadronic
tensor
\begin{equation}
  R_L(q,\omega)= W^{00},     \kern 1cm
  R_T(q,\omega)=W^{11}+W^{22}   .
  \label{eq:responses}
\end{equation}
The inclusive hadronic tensor, \( W^{\mu\nu} \), is constructed from
the matrix elements of the electromagnetic current operator,
\(\hat{J}^{\mu}(\nq),\) between the initial and final hadronic states.
In this article, we mainly focus on independent particle models where
the ground state is approximated by a Slater determinant of single
particle wave functions.  In particular, we examine the Fermi gas (FG)
model, where the single particle states are plane waves
$\psi(\nr)={\rm e}^{i\np\cdot\nr}/\sqrt{V}$, and $V\rightarrow
\infty$ is the volume of the system.  We only consider the final
states that are 1p1h excitations, obtained by promoting a state above
the Fermi level. Other contributions to the response, such as those
due to multinucleon emission or pion emission, are out of the scope
of this work. The 1p1h hadronic tensor in the FG model is
\begin{eqnarray}
W^{\mu\nu}_{1p1h}&=& \sum_{ph}
\left\langle
ph^{-1} \right|\hat{J}^{\mu}(\nq) |\left. F \right\rangle^{*}
\left\langle
ph^{-1} \right|\hat{J}^{\nu}(\nq) |\left. F \right\rangle 
\nonumber
\\ 
&\times& \delta(E_{p}-E_{h}-\omega)
\theta(p-k_F)\theta(k_F-h)
\label{hadronic}
\end{eqnarray}
where $|p\rangle \equiv |\np s_p t_p\rangle$ and $|h\rangle \equiv
|\nh s_h t_h\rangle$ are plane wave states for particles and holes,
respectively, and $|F\rangle$ is the FG ground state with all momenta
occupied below the Fermi momentum $k_F$. 
The delta function ensures energy
conservation in the reaction.
In the  thermodynamic limit, 
$V \rightarrow \infty$, the above sums are transformed into 
 integrals:
\begin{equation}
\sum_{h}
\rightarrow 
V \sum_{s_ht_h}\int \frac{d^3h}{(2\pi)^3}.
\end{equation}

We expand the electromagnetic current as the sum of one-body (1b) plus
two-body (2b) currents
\begin{equation}
\hat{J}^\mu(\nq) = 
\hat{J}^\mu_{1b}(\nq) 
+\hat{J}^\mu_{2b}(\nq),
\end{equation}
Where \(\hat{J}_{1b}\) is the usual electromagnetic
current of the nucleon, while \(\hat{J}_{2b}\) denotes two-body
meson-exchange currents (MEC). 
The matrix element of these operators between the nuclear ground state and 
a 1p1h excitation are given by
\begin{equation}
\left\langle ph^{-1} \right|\hat{J}_{1b}^{\mu} |\left. F \right\rangle
=
\left\langle p \right|\hat{J}_{1b}^{\mu} |\left. h \right\rangle,
\end{equation}
\begin{equation}
\left\langle ph^{-1} \right|\hat{J}_{2b}^{\mu} |\left. F \right\rangle 
=
\sum_{k<k_F}\left[
\left\langle pk \right|\hat{J}_{2b}^{\mu} |\left. hk \right\rangle 
- \left\langle pk \right|\hat{J}_{2b}^{\mu} |\left. kh \right\rangle
\right].
\label{melement}
\end{equation}
The antisymmetry of the total $A$-body wave function implies that the
matrix element of the MEC in Eq. (\ref{melement}) is the sum of a
direct part minus an exchange part, and there is a sum over spectator
states, $k<k_F$,   in the Fermi gas, 
Note that the spectator nucleon,
$|k\rangle = |ks_kt_k\rangle$, does not change its state nor receive
any excitation energy or momentum.

The  elementary matrix elements of the 1b and 2b currents
between plane waves states can be written as:
 \begin{eqnarray}
 \langle p |\hat{J}_{1b}^{\mu} | h\rangle 
&=&
  \frac{(2\pi)^{3}}{V}\delta^{3}(\nq+\nh-\np)
j_{1b}^{\mu}(\np,\nh), 
\label{OBmatrix}
\\
\langle p'_{1}p'_{2}|\hat{J}_{2b}^{\mu}|p_{1}p_{2}\rangle
&=&
\frac{(2\pi)^{3}}{V^{2}}\delta^{3}(\np_1+\np_2+\nq-\np'_1-\np'_2)
\label{TBmatrix}
\nonumber\\
&&
{}\times
j_{2b}^{\mu}(\np'_1,\np'_2,\np_1,\np_2) , 
\label{two-body-matrix}
\end{eqnarray}
where the Dirac deltas arise from momentum conservation.  The current
functions $j^\mu_{1b}(\np,\nh)$ and \(j_{2b}^{\mu}(\np'_1, \np'_2,
\np_1, \np_2)\) implicity depend on spin and isospin indices.

If we define the function
\begin{eqnarray}
j_{2b}^{\mu}(\np,\nh) 
\equiv 
\frac{1}{V}
\sum_{k<k_F}
\left[ j_{2b}^{\mu}(\np,\nk,\nh,\nk)-j_{2b}^{\mu}(\np,\nk,\nk,\nh)\right] ,
\nonumber\\
\label{effectiveOB}
\end{eqnarray}
then  the MEC matrix element can be written in the same
way as that of a one-body operator:
 \begin{eqnarray}
\left\langle ph^{-1} \right|\hat{J}_{2b}^{\mu} |\left. F \right\rangle 
=
\frac{(2\pi)^{3}}{V}\delta^{3}(\nq+\nh-\np)
j_{2b}^{\mu}(\np,\nh).
\nonumber \\
\end{eqnarray}
Therefore the transition matrix element of the total
current between the ground state and the 1p1h state is
 \begin{equation}
\left\langle ph^{-1} \right|\hat{J}^{\mu} |\left. F \right\rangle 
=
  \frac{(2\pi)^{3}}{V}\delta^{3}(\nq+\nh-\np)
j^{\mu}(\np,\nh), 
\label{total}
\end{equation}
whith and effective one-body current for the 1p1h excitation
\begin{equation}
j^{\mu}(\np,\nh)= j_{1b}^{\mu}(\np,\nh)+ j_{2b}^{\mu}(\np,\nh). 
\end{equation}
By inserting (\ref{total}) into Eq. (\ref{hadronic}), taking the
thermodynamic limit, and integrating over the particle momentum, $\np$,
 using the Dirac delta of momentum,
 the hadronic tensor of the Fermi gas can be written as
\begin{eqnarray}
W^{\mu\nu}
&=&
\frac{V}{(2\pi)^{3}}
\sum_{t_h}
\int d^3h\delta(E_{p}-E_{h}-\omega)
2 w^{\mu\nu}(\np,\nh) \nonumber \\
&\times&
\theta(p-k_{F})\theta(k_{F}-h),
\label{integralw}
\end{eqnarray}  
where  \(\np = \nh + \nq\).
This is exactly the same
formula as with the one-body current, with the difference that the
current now includes the contribution of the MEC.
The  function $w^{\mu\nu}$ is the effective single-nucleon hadronic tensor 
in the transition $h\rightarrow p$
\begin{eqnarray}
w^{\mu\nu}(\np,\nh)=\frac{1}{2}\sum_{s_ps_h}
j^\mu(\np,\nh)^*j^\nu(\np,\nh) .
\end{eqnarray}
This single-nucleon tensor implicitly refers either to a proton or a
neutron, and the summation over isospin \( t_h \) in
Eq. (\ref{integralw}) corresponds to summing the responses of protons,
  $t_h=\frac12$, and neutrons, $t_h=-\frac12$, together.

The effective single-nucleon tensor incorporates the contribution of
the MEC, implying there is interference between 1b and 2b currents.
In fact, the response functions only involve diagonal elements of the
hadronic tensor, and the transverse component \(\mu\mu\) (with
$\mu=1,2$) can be expanded as
\begin{eqnarray}
w^{\mu\mu}(\np,\nh)
&=& \frac{1}{2}\sum_{s_ps_h} |j^\mu_{1b}(\np,\nh)+j^\mu_{2b}(\np,\nh)|^2
\nonumber\\
&=&
  w^{\mu\mu}_{1b}+ w^{\mu\mu}_{1b2b}+w^{\mu\mu}_{2b},
\label{single-nucleon}
\end{eqnarray}
where 
\begin{eqnarray}
w^{\mu\mu}_{1b}   & = & \frac{1}{2}\sum |j^\mu_{1b}|^2, \\
w^{\mu\mu}_{1b2b} & = & \mbox{Re}\sum (j^\mu_{1b})^*j^\mu_{2b}, \label{w1b2b}\\
w^{\mu\mu}_{2b}   & = & \frac{1}{2}\sum |j^\mu_{2b}|^2.
\end{eqnarray}
The first term, \(w^{\mu\mu}_{1b}\),
is the tensor corresponding to the one-body current alone, \(w^{\mu\mu}_{12}\)
is the interference between 1b and 2b
currents, and \(w^{\mu\mu}_{2b}\) represents the contribution of the two-body
current alone.
The 1b part is the leading contribution in the quasielastic peak,
 while the dominant
contribution of the MEC corresponds to the interference with the
one-body current   \cite{Ama94a,Ama03},
 being the pure contribution of the two-body current  generally smaller.

Therefore the transverse response function is the sum of a response induced by
the one-body current, plus an interference term between the 1b and the
2b currents, plus a response due solely to the MEC. Typically, the
interference term dominates over the pure MEC contribution, as the MEC
represent a small perturbation relative to the one-body current;  
\begin{equation}
R^T= R^T_{1b}+ R^T_{1b2b} + R^T_{2b}.
\end{equation} 
In this work, we focus on the transverse interference response between
1b and 2b currents $R^T_{1b2b}$. We will demonstrate
the low momentum results in the non-relativistic limit
(the contribution of the MEC in the longitudinal channel is of higher
order in the non-relativistic limit and can be neglected). 

 In the non-relativistic limit, the 1b current is the sum of
 magnetization and convection currents:
\begin{eqnarray}
  \nj_{1b}(\np,\nh)
&=& 
  \nj_{M}(\np,\nh)+  \nj_{C}(\np,\nh), 
\\
  \nj_{M}(\np,\nh)
&=&
-\delta_{t_pt_h}\frac{G_M^h}{2m_N}i\nq\times\nsigma_{s_ps_h},
\label{magnetization}
\\
  \nj_C(\np,\nh)
&=&
\delta_{t_pt_h}\delta_{s_ps_h}
\frac{G_E^h}{m_N}(\nh+\frac{\nq}{2}).
\label{convection}
\end{eqnarray}
with $\nq=\np-\nh$ by momentum conservation.  Here $G_M^h$ ($G_E^h$)
is the magnetic (electric) form factor of the nucleon with isospin
$t_h$.  In the quasielastic peak the convection current contribution
is much smaller that the magnetization and can be neglected.

\subsection{Meson exchange currents}

The starting point in this work is the relativistic MEC model from
reference \cite{Rui17}, derived from the pion production model of
 \cite{Her07}, that follows from the Lagrangian given in Appendix \ref{Appendix0}.
 This model has been used to describe the 2p2h
response in both electron and neutrino scattering \cite{Ama20}, and
has been implemented in the neutrino event generator code GENIE
\cite{Dol19}.  We begin with the relativistic MEC model and
systematically derive its non-relativistic reduction. This approach
ensures consistency beteen  relativistic and non-relativistic
currents. Additionally, it
allows us to connect our results with previous studies that employed
non-relativistic calculations, facilitating a direct comparison and
validation of our approach.  The non-relativistic MEC will serve as
the basis for deriving the low-momentum results  
for the 1p1h transverse response in
the low momentum transfer regime, allowing us to connect with previous
works using non-relativistic models. We consider low momentum transfer
to be around \( q \sim 500 \) MeV/c or lower.

\begin{figure}[htb]
\centering
\includegraphics[width=8cm,bb=120 440 495 700]{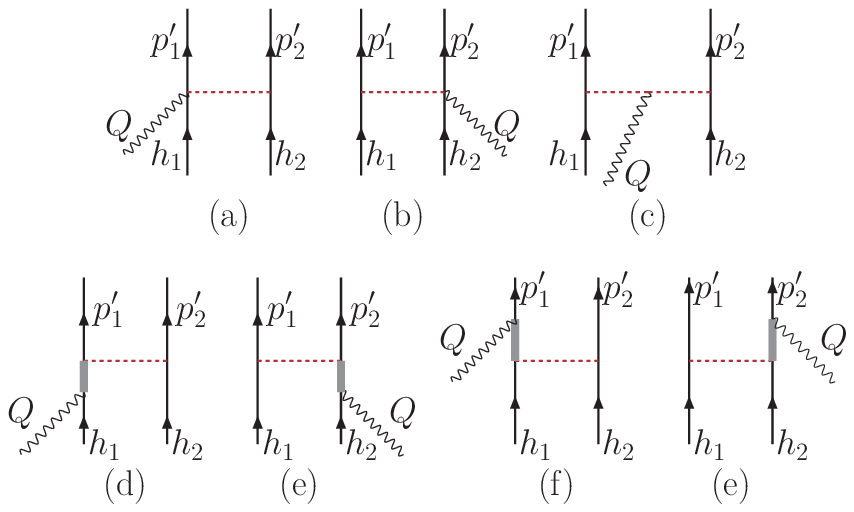}
\caption{Feynman diagrams for the MEC model used in 
this work: seagull (a,b), pion-in-flight (c) and $\Delta$ (d,e,f,g).}
\label{feynman}
\end{figure}

The relativistic MEC are obtained as the sum of the diagrams depicted
in Fig. 1, which include seagull (s), pion-in-flight ($\pi$) ,
and \(\Delta\) isobar currents.
\begin{equation}
j_{2b}^\mu(\np'_1,\np'_2,\np_1,\np_2) 
=  j^{\mu}_{s}+  j^{\mu}_{\pi}+  j^{\mu}_{\Delta},
\end{equation}
where the $\Delta$ current is the sum of forward and backward terms
\begin{equation}
 j^{\mu}_{\Delta}= j^{\mu}_{\Delta F}+ j^{\mu}_{\Delta B}.
\end{equation}
The specific treatment of the $\Delta$ current is model-dependent, and
various versions exist due to uncertainty in the off-shell properties
of the $\Delta$ and its interaction with the medium. Other models used
for relativistic MEC include those described in
Refs. \cite{Pac03,Ama03,Pas95}, although the 1p1h transverse response
 does not differ significantly between them and the model
presented here. In particular, a recent calculation \cite{Cas23}
of 1p1h responses with the present MEC model was compared with 
those of the model of
Ref. \cite{Ama03}, and very similar results were found.

The MEC matrix elements in the present model are given by 
\begin{widetext}
\begin{eqnarray} \label{2p2h}
  j^{\mu}_{sea}
&=&
i[\ntau^{(1)} \times \ntau^{(2)}]_z
\frac{f^{2}}{m_{\pi}^{2}}
V(1',1)
F_{\pi NN}(k_{1}^{2})
\bar{u}_{s'_2}(p'_2)F_{1}^{V}\gamma^{5}\gamma^{\mu}u_{s_2}(p_{2}) 
+ (1 \leftrightarrow 2)
\\
  j^{\mu}_{\pi}
&=&
i[\ntau^{(1)} \times \ntau^{(2)}]_z
\frac{f^{2}}{m_{\pi}^{2}}F_{1}^{V}
V(1',1)
V(2',2)
(k_{1}^{\mu}-k_{2}^{\mu}) 
\\  
  j^{\mu}_{\Delta F}
&=&
U_F(1,2)
\frac{ff^{*}}{m_{\pi}^{2}}
V(2',2)
F_{\pi N \Delta}(k_{2}^{2})
\bar{u}_{s'_1}(p'_{1})k_{2}^{\alpha}G_{\alpha\beta}(p_{1}+Q)
\Gamma^{\beta\mu}(Q)u_{s_1}(p_{1}) 
+ (1 \leftrightarrow 2)
\label{delta1}
\\
  j^{\mu}_{\Delta B}
&=&
U_B(1,2)
\frac{ff^{*}}{m_{\pi}^{2}}
V(2',2)
F_{\pi N \Delta}(k_{2}^{2})
\bar{u}_{s'_1}(p'_{1})k_{2}^{\beta}
\Gamma^{\alpha\mu}(-Q)G_{\alpha\beta}(p'_{1}-Q)u_{s_1}(p_{1}) 
+ (1 \leftrightarrow 2).  
\label{delta2}
\end{eqnarray}
\end{widetext}
Here a matrix element between initial and final isospin states is implicit, 
i.e., $\langle t'_1t'_2| j^\mu | t_1t_2\rangle$,
although we do not explicitly write this to simplify the notation.
The  symbols appearing in the MEC matrix elements are the following: 

\begin{itemize}

\item 
The $\pi N$ coupling constant, $f^2=1$, and the $\pi N \Delta$
coupling constant, $f^*=2.13 f$.

\item 
The spinors \(u(p)\) are the solutions of the Dirac equation with
momentum $\np$.

\item
The four-vector
$k_{i}^{\mu}=p'_{i}{}^{\mu}-p_{i}^{\mu}$ is the momentum
transferred to the nucleon $i=1,2$.

\item  $\ntau^{(i)}$ is the isospin operator of  nucleon $i$.

\item $F_1^V(Q^2) =F_1^p-F_1^n$ is the isovector form factor of the nucleon.

\item The following function of spin and momentum
is common to all the currents,
 including a $\pi NN$ form factor, and the
pion propagator 
\begin{equation}
  V(1',1) \equiv 
 F_{\pi NN}(k_{1}^{2})
\frac{\bar{u}_{s'_1}(p'_{1})\gamma^{5}\kbar_{1}u_{s_1}(p_{1})}{k_1^2-m_{\pi}^2}.
\label{V-function}
\end{equation}

\item 
The $\pi$NN and $\pi$N$\Delta$ form factors at the pion absorption/emission 
vertices are \cite{Alb84,Som78} 
\begin{equation}
  F_{\pi NN}(k)=   F_{\pi N\Delta}(k)
=\frac{\Lambda^{2}-m_{\pi}^{2}}{\Lambda^{2}-k^{2}},
\label{pinnff}
\end{equation}
where the parameter \(\Lambda\)  
is typically chosen to be around 1300 MeV \cite{Mac87}.

\end{itemize}

The forward $\Delta$ current corresponds to processes where the
$\Delta$ resonance is produced and then decays back to a nucleon,
while the backward $\Delta$ current involves the exchange of a pion,
leading to the creation of a $\Delta$ resonance in the intermediate
state. The charge dependence of these processes is embedded in the
isospin operators
\begin{eqnarray}
 U_{F}(1,2) 
&=&
\sqrt{\frac{3}{2}}
\sum_{i=1}^{3}
T_{i}^{(1)}T_{z}^{(1)\dagger}\tau_{i}^{(2)},
\label{uf}\\ 
 U_{B}(1,2) 
&=&
\sqrt{\frac{3}{2}}
\sum_{i=1}^{3}
T_{z}^{(1)}T_{i}^{(1)\dagger}\tau_{i}^{(2)},
\label{ub}
\end{eqnarray}
where $T^{\dagger}_i$ are the Cartesian coordinates of the 
$\frac{1}{2} \rightarrow \frac{3}{2}$
transition isospin operator, defined by \cite{Eri88}  
\begin{equation}
\textstyle
\langle \frac32 t_\Delta | T^\dagger_\mu | \frac12 t \rangle
= \langle \frac12 t 1 \mu | \frac32 t_\Delta \rangle
\end{equation}  
$T^\dagger_\mu$ 
 being the spherical components of the vector
$\vec{T}^\dagger$.  

The following $\gamma N\rightarrow \Delta$ transition vertex
\cite{Lle72,Her07} is used as the leading contribution for low momentum
transfer
\begin{equation} \label{gammabetamu}
  \Gamma^{\beta\mu}(Q)=
\frac{C_3^V}{m_N}
(g^{\beta\mu}\Qbar-Q^{\beta}\gamma^{\mu})\gamma_5.
\end{equation}
We use the $\Delta$ vector form factor \cite{Her07}:
\begin{equation}
  C_{3}^{V}(Q^{2})
=\frac{2.13}{(1-\frac{Q^{2}}{M_{V}^{2}})^{2}}
\frac{1}{1-\frac{Q^{2}}{4M_{V}^{2}}} .
\label{C3}
\end{equation}
The $\Delta$ propagator is
\begin{equation}
  G_{\alpha\beta}(P)=
\frac{{\cal P}_{\alpha\beta}(P)}{
P^{2}-M_{\Delta}^{2}+iM_{\Delta}\Gamma(P^{2})+\frac{\Gamma(P^{2})^{2}}{4}}
\end{equation} 
where $M_{\Delta}$ and $\Gamma$ are the $\Delta$ mass and width
respectively. The projector ${\cal P}_{\alpha\beta}(P)$ over spin-3/2 
is
\begin{eqnarray}
{\cal  P}_{\alpha\beta}(P)
&=&
-(\Pbar+M_{\Delta})
\nonumber\\
&&
\kern -1cm
 \times
\left[
g_{\alpha\beta}-\frac{\gamma_{\alpha}\gamma_{\beta}}{3}-\frac{2P_{\alpha}P_{\beta}}{3M_{\Delta}^{2}}+\frac{P_{\alpha}\gamma_{\beta}-P_{\beta}\gamma_{\alpha}}{3M_{\Delta}}
\right] .
\end{eqnarray}  
The 
$\Delta$ width $\Gamma(P^{2})$ is given by
\begin{equation}
  \Gamma(P^{2})=\Gamma_{0}\frac{m_{\Delta}}{\sqrt{P^{2}}}
\left(\frac{p_{\pi}}{p^{res}_{\pi}}\right)^{3}.
\label{width}
\end{equation}
where $\Gamma_{0}=120$ MeV is the $\Delta$ width at rest, $p_{\pi}$ is
the momentum of the final pion in the $\Delta$ decay, and
$p^{res}_{\pi}$ is its value at resonance ($P^2=m_\Delta^2$).

Some details about the treatment of the \(\Delta\)
propagator, which are model-dependent and not entirely
established, are not relevant in the non-relativistic
limit considered here but can be significant when applying a
relativistic model or making relativistic corrections.  The width
\(\Gamma_{\Delta}\) corresponds to the \(\Delta\) in vacuum, and it is
expected to be slightly different in the medium depending on the
kinematics.
Various alternative approximations to the \(\Delta\) propagator have
been proposed. However, in the case of the quasielastic peak, the
typical kinematics are of the order of 1 GeV, and these issues are not
expected to be relevant. They are overshadowed by other more
significant nuclear effects that dominate in this energy regime.

In the non-relativistic limit, which we focus on in this work, these
model-dependent details of the \(\Delta\) propagator are not
critical. The primary concern is ensuring that our non-relativistic
reduction of the MEC operators is consistent with the standard
non-relativistic MEC operators used in previous studies. This
consistency guarantees that our results are reliable and can be
reproduced using the same theoretical framework.

\subsection{Non relativistic MEC}

In this section, we derive in detail the non-relativistic reduction of
the meson exchange currents (MEC). This treatment is deliberately
didactic, as mentioned, to ensure that our results are easily
reproducible. We aim to leave no ambiguous or unclear steps in the
derivation process. We will take the static limit in which the
momenta of the initial and final nucleons are very small. 

In the non-relativistic limit, the lower component of the Dirac
spinors is neglected, and the $4\times 4$ Dirac matrices reduce to $2\times 2$  Pauli 
matrices
acting on the upper components. Additionally, we will only consider
the non-relativistic reduction of the transverse current, i.e.,
\(J^i\) for \(i=1,2\), as this is the dominant contribution in this
limit. The contribution of MEC to the longitudinal response is
neglected as it is of higher order in the non-relativistic limit.

The non-relativistic approximation is further justified by numerical
calculations. Although it is not the goal of this article to perform
an exhaustive comparison of the longitudinal response \(R_L\) in the
fully relativistic model, numerical verification shows that the
contribution of MEC to \(R_L\) is indeed negligible compared to its
contribution to the transverse response.

To achieve this reduction, we apply the following rules:
\begin{eqnarray}
\gamma^0 \longrightarrow 1, &
\gamma^i \longrightarrow 0, &
\gamma_5\gamma^0 \longrightarrow 0, 
\label{gammas1}\\
\gamma_5\gamma^i \longrightarrow -\sigma_i, &
\gamma^i\gamma^j \longrightarrow -\sigma_i\sigma_j, &
\gamma^0\gamma^j \longrightarrow 0. 
\label{gammas2}
\end{eqnarray}
For a nucleon momentum: 
\begin{eqnarray}
 p^\mu \longrightarrow (m_N,p^i), &&
\pbar \longrightarrow p_0 
\end{eqnarray}
For the momentum transfer to nucleon $i$:
\begin{eqnarray}
k^\mu \longrightarrow (0,k^i), &&
\gamma_5\kbar \longrightarrow \nk\cdot\nsigma.
\end{eqnarray}
To simplify the writing at this stage, we do not explicitly include
the strong form factors. These can be applied later on to the
non-relativistic currents. Most of the results will be obtained with
these form factors set to one, and we will see that the effect of
including them is small for the low momentum transfer values
considered here. Additionally, their inclusion does not alter the sign of  the interference. 

 The $V$-function of Eq. (\ref{V-function}) is directly obtained in this limit
\begin{equation}
  V(1',1) \longrightarrow 
-\frac{\nk_1\cdot\nsigma^{(1)}}{\nk_1^2+m_{\pi}^2}
\label{Vnorel}
\end{equation}
where a matrix element between initial and final spin states is
understood, i.e. $\langle s'_1| \cdot | s_1\rangle$, but is not
explicitly written for simplicity. The spin states $|s\rangle$ are non
relativistic, two-component spinors, corresponding to the upper
component of the Dirac spinors.

\subsubsection{Seagull current}

In the seagull current, we start by using a notation to separate the
isospin part, which does not change in the non-relativistic limit,
from the spin-momentum part, which does change.
\begin{eqnarray} 
  j^{\mu}_{s}
&=&
i[\ntau^{(1)} \times \ntau^{(2)}]_z
(K^\mu_{s}-L^\mu_{s})
\\
&=&
i[\ntau^{(1)} \times \ntau^{(2)}]_z
j^\mu_{s3}.
\end{eqnarray}
Here the auxiliar functions 
$K^\mu_{s}$, $L^\mu_{s}$ and $j^\mu_{s3}$ are independent on isospin
and are defined by 
\begin{eqnarray}
K^\mu_{s}(1',2',1,2)&=&
\frac{f^{2}}{m_{\pi}^{2}}
V(1',1)
\bar{u}(2')F_{1}^{V}\gamma^{5}\gamma^{\mu}u(2) 
\\
L^\mu_{s}(1',2',1,2)&=&K^\mu_{s}(2',1',2,1)
\\
j^\mu_{s3} &=& K^\mu_{s}-L^\mu_{s},
\end{eqnarray}
where in the $(1\leftrightarrow 2)$ term we have used the property
\begin{equation}
[\ntau^{(2)} \times \ntau^{(1)}] =-[\ntau^{(1)} \times \ntau^{(2)}].
\end{equation}
The non-relativistic reduction is directly obtained
using Eqs. (\ref{gammas2}) and (\ref{Vnorel})
\begin{eqnarray}
K^i_{s} \rightarrow 
\frac{f^{2}}{m_{\pi}^{2}}F_1^V
\frac{\nk_1\cdot\nsigma^{(1)}}{\nk_1^2+m_{\pi}^2}
\sigma_i^{(2)}
\end{eqnarray}
where, as before, a matrix element between two-components initial and
final spin states is understood, $\langle
s'_1s'_2|\cdot|s_1s_2\rangle$.  Therefore the spin-momentum part of
the seagull current $j_{s3}^i$ in the non-relativistic limit becomes
\begin{equation}
\nj_{s3} \rightarrow 
\frac{f^{2}}{m_{\pi}^{2}}F_1^V
\left(
\frac{\nk_1\cdot\nsigma^{(1)}}{\nk_1^2+m_{\pi}^2}
\nsigma^{(2)}
-\frac{\nk_2\cdot\nsigma^{(2)}}{\nk_2^2+m_{\pi}^2}
\nsigma^{(1)}
\right).
\label{seagull}
\end{equation}

\subsubsection{Pionic current}

The pion-in-fligh  current can be   written similarly to the seagull current as
\begin{equation} 
  j^{\mu}_{\pi}=
i[\ntau^{(1)} \times \ntau^{(2)}]_z
j^\mu_{\pi 3} 
\end{equation}
where  
$j^\mu_{\pi 3}$ is independent on isospin
and is defined by 
\begin{equation}
j^\mu_{\pi 3}(1',2',1,2)
=
\frac{f^{2}}{m_{\pi}^{2}}
F_{1}^{V} V(1',1)V(2',2)(k_1^\mu-k_2^\mu).
\end{equation}
The non-relativistic reduction of the seagull current is directly obtained
from Eq. (\ref{Vnorel})
\begin{equation}
\nj_{\pi 3} \rightarrow 
\frac{f^{2}}{m_{\pi}^{2}}F_1^V
\frac{\nk_1\cdot\nsigma^{(1)}}{\nk_1^2+m_{\pi}^2}
\frac{\nk_2\cdot\nsigma^{(2)}}{\nk_2^2+m_{\pi}^2}
(\nk_1-\nk_2).
\label{pionic}
\end{equation}
It is
straightforward to verify that
the non-relativistic reduction of the seagull plus pionic currents 
coincide with the usual expressions
found in the literature \cite{Ris89}.

\subsubsection{$\Delta$ current}

The non-relativistic reduction of the $\Delta$ current is somewhat more involved due
to its more complex spin and isospin structure. This complexity arises
from the transition between nucleon and $\Delta$ states, and the $\Delta$
propagator, which introduces additional terms that must be
carefully managed during the non-relativistic limit process.
In order to simplify the non-relativistic reduction, 
it is convenient to write the $\Delta$ current in an abbreviated
form 
\begin{eqnarray}
  j^{\mu}_{\Delta F}
&=&
U_F K_F^\mu 
+ (1 \leftrightarrow 2)
\\
  j^{\mu}_{\Delta B}
&=&
U_B K_B^\mu 
+ (1 \leftrightarrow 2)
\end{eqnarray}
where a matrix element is assumed to be taken between the initial and final
isospin states $\langle t'_1 t'_2 |\cdot |t_1t_2\rangle$, 
but we do not explicitly write this to simplify the
notation.

The functions $K^\mu_F$ and $K^\mu_B$ are independent on isospin
and can be written as
\begin{eqnarray}
  K^{\mu}_{ F}
&=&
\frac{ff^{*}}{m_{\pi}^{2}}
V(2',2)
A^\mu,
\\
  K^{\mu}_{B}
&=&
\frac{ff^{*}}{m_{\pi}^{2}}
V(2',2)
B^\mu.
\end{eqnarray}
Finally $A^\mu$ and $B^\mu$ are defined as
\begin{eqnarray}
A^\mu
&=&
\bar{u}(1')k_{2}^{\alpha}G_{\alpha\beta}(p_{1}+Q)
\Gamma^{\beta\mu}(Q)u(1) 
\label{amu}\\
B^\mu
&=&
\bar{u}(1')k_{2}^{\beta}
\Gamma^{\alpha\mu}(-Q)G_{\alpha\beta}(p'_{1}-Q)u(1). 
\label{bmu}
\end{eqnarray}
The non-relativistic reduction of the spatial components of the $\Delta$
current requires the non-relativistic reduction of the components $A^i$
and $B^i$. This detailed reduction process is carried out in Appendix \ref{AppendixA}.
The result is
\begin{eqnarray}
\nA &\rightarrow & 
g_\Delta
\nq\times\left[\frac23 i \nk_2 - \frac13 \nk_2\times\nsigma^{(1)}\right]
\label{Anonrel}\\
\nB &\rightarrow & 
g_\Delta
\nq\times\left[\frac23 i \nk_2 + \frac13 \nk_2\times\nsigma^{(1)}\right],
\label{Bnonrel}
\end{eqnarray}
where we have defined the constant
\begin{equation}
g_\Delta \equiv \frac{c_3^V}{m_N}\frac{1}{m_\Delta-m_N}.
\end{equation}
Therefore
\begin{eqnarray}
\nK_F &\rightarrow & 
-\frac{ff^*}{m_\pi^2}
\frac{\nk_2\cdot\nsigma^{(2)}}{\nk_2^2+m_{\pi}^2}\nA
\\
\nK_B &\rightarrow & 
-\frac{ff^*}{m_\pi^2}
\frac{\nk_2\cdot\nsigma^{(2)}}{\nk_2^2+m_{\pi}^2}\nB.
\end{eqnarray}
As before, a matrix element is assumed to be taken between the initial and final
spin states, $\langle s'_1 s'_2 |\cdot |s_1s_2\rangle$, 
but we do not explicitly write this to simplify the
notation.

Using the result
\begin{equation} \label{titj}
T_{i}T^{\dagger}_{j}=
\frac23\delta_{ij}
-\frac{i}{3}\epsilon_{ijk}\tau_{k}
=\delta_{ij}-\frac{1}{3}\tau_{i}\tau_{j}
\end{equation}

the forward and backward isospin operators can be written as
\begin{eqnarray}
  U_F(1,2)
&=&
\frac{1}{\sqrt{6}}
\left(2\tau_{z}^{(2)}
      -i[\tau^{(1)}\times\tau^{(2)}]_{z}
\right)
\\
 U_B(1,2)
&=&
\frac{1}{\sqrt{6}}
\left(2\tau_{z}^{(2)}
      +i[\tau^{(1)}\times\tau^{(2)}]_{z}
\right).
\end{eqnarray}
Hence the $\Delta$ current can be written as
\begin{eqnarray}
\nj_\Delta 
&=&
\nj_{\Delta F} +\nj_{\Delta B}
\nonumber\\
&=&
\frac{2}{\sqrt{6}}
\tau_{z}^{(2)}
[\nK_F+\nK_B]
\label{jdeltanr}\\
&&
+\frac{1}{\sqrt{6}}
i[\tau^{(1)}\times\tau^{(2)}]_{z}
[\nK_B-\nK_F]
+ (1\leftrightarrow 2).
\nonumber
\end{eqnarray} 
We see that the isospin operators in the $\Delta$ current, $\tau^{(2)}_z$ 
and $[\ntau^{(1)}\times \ntau^{(2)}]_z$, 
are multiplied by the sum and the difference between the backward and forward 
functions, \( \nK_B \) and \( \nK_F \), given in the non-relativistic limit by
\begin{eqnarray}
\nK_F+\nK_B 
&=&
-\frac{ff^*}{m_\pi^2}
\frac{\nk_2\cdot\nsigma^{(2)}}{\nk_2^2+m_{\pi}^2}(\nA+\nB)
\label{FpB}\\
\nK_B-\nK_F 
&=&
-\frac{ff^*}{m_\pi^2}
\frac{\nk_2\cdot\nsigma^{(2)}}{\nk_2^2+m_{\pi}^2}(\nB-\nA).
\label{BmF}
\end{eqnarray}
From Eqs. (\ref{Anonrel}) and (\ref{Bnonrel}),
\begin{eqnarray}
\nA+\nB &=& \frac43 g_\Delta i \nq\times\nk_2
\\
\nB-\nA &=& \frac 23 g_\Delta \nq\times(\nk_2\times\nsigma^{(1)}).
\end{eqnarray}
Inserting this result into Eqs. (\ref{FpB}) and (\ref{BmF}) we have
\begin{eqnarray}
\nK_F+\nK_B 
&=&
-
\frac43 g_\Delta 
\frac{ff^*}{m_\pi^2}
\frac{\nk_2\cdot\nsigma^{(2)}}{\nk_2^2+m_{\pi}^2}
i \nq\times\nk_2,
\\
\nK_B-\nK_F 
&=&
-\frac 23 g_\Delta 
\frac{ff^*}{m_\pi^2}
\frac{\nk_2\cdot\nsigma^{(2)}}{\nk_2^2+m_{\pi}^2}
\nq\times(\nk_2\times\nsigma^{(1)}).
\nonumber\\
\end{eqnarray}
Using these results in Eq. (\ref{jdeltanr}), it is straightforward to obtain
the following expression for the non-relativistic $\Delta$ current
\begin{widetext}
\begin{eqnarray}
\nj_\Delta 
&=&
i \sqrt{ \frac32 } \frac29 
\frac{ff^*}{m_\pi^2}
\frac{c_3^V}{m_N}\frac{1}{m_\Delta-m_N}
\left\{
\frac{\nk_2\cdot\nsigma^{(2)}}{\nk_2^2+m_{\pi}^2}
\left[
4\tau_{z}^{(2)}\nk_2+
[\tau^{(1)}\times\tau^{(2)}]_{z}
\nk_2\times\nsigma^{(1)}
\right]
\right.
\nonumber\\
&&
\kern 4cm 
\left. \mbox{}+
\frac{\nk_1\cdot\nsigma^{(1)}}{\nk_1^2+m_{\pi}^2}
\left[
4\tau_{z}^{(1)}\nk_1-
[\tau^{(1)}\times\tau^{(2)}]_{z}
\nk_1\times\nsigma^{(2)}
\right]
\right\}
\times\nq.
\label{deltafinal}
\end{eqnarray} 
\end{widetext}
One can, in fact, verify that this expression coincides with the
$\Delta$ current appearing in the literature, particularly the
expression given in Refs. \cite{Ris89,Sch89}, that we use as reference,
except for the precise values of the coupling constants and form
factors. This assures us that the relativistic and non-relativistic
calculations in the low-momentum and low-energy limit should coincide
if identical values of coupling and form factors are used.

\subsection{MEC effective one-body transition currents}

With the non-relativistic MEC current obtained in the last section,
 the effective one-body transition current
\(\nj_{2b}(\np,\nh)\) in the Fermi gas is obtained by summing over the
spin, isospin, and integrating over the momentum \(k\) of the
spectator nucleon.  At leading order, only the spatial part of the MEC
survives, affecting solely the transverse response, which is
perpendicular to the transferred momentum \(\nq\).  From
Eq. (\ref{effectiveOB}) in the $V\rightarrow \infty$ this current is
\begin{equation}
\nj_{2b}(\np,\nh) 
=
\int\frac{d^3k}{(2\pi)^3}
\sum_{t_ks_k}
\left[ \nj_{2b}(p,k,h,k)-\nj_{2b}(p,k,k,h)\right] .
\label{effective}
\end{equation}

\subsubsection{Sum over isospin}

 We begin by showing that the direct term in Eq. (\ref{effective})
 is zero.  Previously, we note that the MEC can be expanded in terms
 of the isospin operators \(\tau^{(1)}_z\) , \(\tau^{(2)}_z\) and
 \([\ntau^{(1)} \times \ntau^{(2)}]_z\)
\begin{equation}
\nj_{2b}= \tau^{(1)}_z \nj_1+ \tau^{(2)}_z \nj_2 + 
i[ \ntau^{(1)} \times \ntau^{(2)}]_z \ \nj_3
\end{equation}
where $\nj_1$, $\nj_2$, $\nj_3$ are independent on isospin.

\begin{figure}[ht]
  \centering
\includegraphics[width=8cm,bb=110 440 500 695]{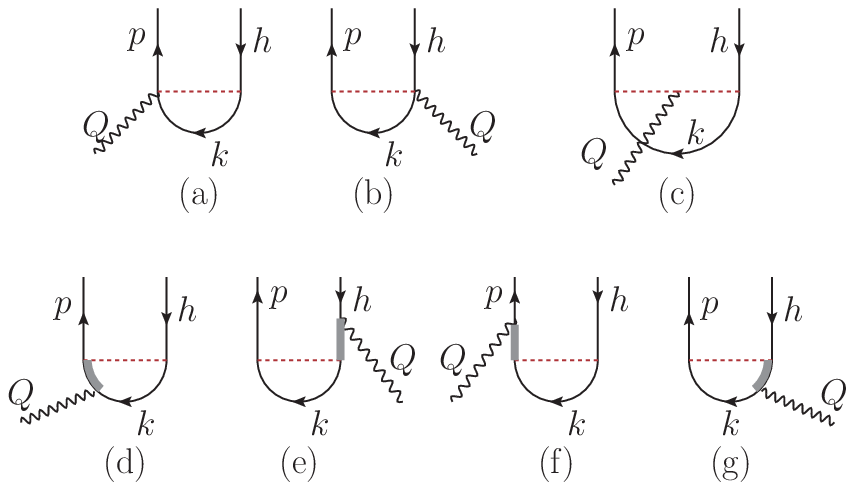}
\caption{Many-body diagrams for the 1p1h MEC matrix elements}
\label{feynman2}
\end{figure}

We first perform  the sum over isospin index $t_k$. 
The direct term is
\begin{eqnarray}
\sum_{t_k}\nj_{2b}(p,k,h,k)
&=&
\sum_{t_k} 
\langle t_pt_k|
\tau^{(1)}_z \nj_1+ \tau^{(2)}_z \nj_2 + 
\nonumber\\
&&+i[\ntau^{(1)} \times \ntau^{(2)}]_z\ \nj_3
| t_ht_k\rangle
\nonumber\\
&=& \delta_{t_pt_h}4t_h \nj_1(p,k,h,k),
\end{eqnarray}
where we have used the elementary isospin sums performed in Appendix \ref{AppendixB}, Eqs. (\ref{iso1}--\ref{iso3})
Therefore the direct term in the matrix element (\ref{effective}) 
is proportional to
\(\nj_1(\np,\nk,\nh,\nk)\), which turns out to be zero. Indeed, \(\nj_1\)
can be obtained from equation (\ref{deltafinal}) as
\begin{equation}
\nj_1 
=
iC_\Delta
\frac{\nk_1\cdot\nsigma^{(1)}}{\nk_1^2+m_{\pi}^2}
4\nk_1\times\nq,
\end{equation} 
with 
\begin{equation}  \label{cdelta}
C_\Delta\equiv
 \sqrt{ \frac32 } \frac29 
\frac{ff^*}{m_\pi^2}
\frac{C_3^V}{m_N}\frac{1}{m_\Delta-m_N}
\end{equation}
and $\nk_1=\np-\nh=\nq$. Therefore 
\begin{equation}
\sum_{t_k}\nj_{2b}(p,k,h,k)
=0.
\end{equation}
This results follows because he $\Delta$
current is transverse, i.e., perpendicular to $\nq$.  In the
relativistic case, a similar situation occurs, and the direct term is
zero although the demonstration is more involved. It requires summing
over the spin of \(\nk\) and handling a large number of terms that
involve many \(\gamma\) matrices.

\begin{widetext}

The sum over isospin in the exchange part is obtained using the isospin sums performed in Appendix \ref{AppendixB}, Eqs (\ref{iso4}--\ref{iso6})
\begin{eqnarray}
\sum_{t_k}\nj_{2b}(p,k,k,h)
&=&
\sum_{t_k=\pm 1/2} 
\langle t_pt_k|
\tau^{(1)}_z \nj_1+ \tau^{(2)}_z \nj_2 + 
i[\ntau^{(1)} \times \ntau^{(2)}]_z\ \nj_3
| t_kt_h\rangle
\nonumber\\
&=& \delta_{t_pt_h}2t_h [\nj_1(p,k,k,h)+\nj_2(p,k,k,h)-2\nj_3(p,k,k,h)].
\end{eqnarray}

Here we observe that in symmetric nuclear matter, the direct matrix
element of the MEC vanishes, and only the exchange term
survives in the 1p1h matrix element. 
Therefore, the many-body diagrams that contribute to the MEC
in the 1p1h channel are those shown in Fig. \ref{feynman2}. Next, we proceed to
perform the spin sums for the different terms of the current.

It can be shown that, in fact, two of the four exchange diagrams
contributing to the $\Delta$ current in the 1p1h matrix element are
zero. Specifically, due to the sum over isospin in the forward
$\Delta$ current, the diagram containing the isospin operator
\(U_F(2,1)\) yields zero (diagram (e) of Fig. 2). Similarly, for the
backward $\Delta$ current, the diagram involving the isospin operator
\(U_B(1,2)\) also vanishes (diagram (f) of Fig. 2). Therefore only
diagrams (d), forward,  and (g), backward,
 contribute in the case of the $\Delta$
current. These results are demonstrated in Appendix \ref{AppendixA}.

\subsubsection{Sum over spin}

The resulting 1p1h matrix elements of the 2b current are computed as
\begin{equation}
\nj_{2b}(p,h)=
-\int \frac{d^3k}{(2\pi)^3}
\sum_{t_ks_k}\nj_{2b}(p,k,k,h)
=
\nj_{s}(p,h)+\nj_{\pi}(p,h)+\nj_{\Delta}(p,h),
\end{equation}
where only the exchange part contributes. 
The sums over spin index $s_k$ are performed in Appendix \ref{AppendixC}. The results are the following for the three MEC, seagull, pionic and $\Delta$ currents

\begin{eqnarray}
\nj_s(p,h)
&=&
4t_h\delta_{t_pt_h} 
\frac{f^2}{m_\pi^2}F_1^V
\int \frac{d^3k}{(2\pi)^3}
\left(
\frac{\delta_{s_ps_h}\nk_1+i\nsigma_{ph}\times\nk_1}{\nk_1^2+m_{\pi}^2}
-\frac{\delta_{s_ps_h}\nk_2+i\nk_2\times\nsigma_{ph}}{\nk_2^2+m_{\pi}^2}
\right)
\label{seagullph}
\\
\nj_\pi(p,h)
&=&
4t_h\delta_{t_pt_h} 
\frac{f^2}{m_\pi^2}F_1^V
\int \frac{d^3k}{(2\pi)^3}
\frac{\delta_{s_ps_h}\nk_1\cdot\nk_2
+i(\nk_1\times\nk_2)\cdot\nsigma_{ph}}
{(\nk_1^2+m_{\pi}^2)(\nk_1^2+m_{\pi}^2)}(\nk_1-\nk_2)
\label{pionicph}
\\
\nj_\Delta(p,h)
&=&
4it_h\delta_{t_pt_h}
C_\Delta\nq\times 
\int \frac{d^3k}{(2\pi)^3}
\left(
\frac{\nk_1^2\nsigma_{ph}+(\nsigma_{ph}\cdot\nk_1)\nk_1}
     {\nk_1^2+m_\pi^2}
+\frac{\nk_2^2\nsigma_{ph}+(\nsigma_{ph}\cdot\nk_2)\nk_2}
     {\nk_2^2+m_\pi^2}
\right)
\label{deltaph}
\end{eqnarray}
with
$\nk_1=\np-\nk$ 
and 
$\nk_2=\nk-\nh$.

\end{widetext}

\subsection{Interference between one-body and MEC in the transverse response}

In this section, we give the MEC contribution to the effective
single-nucleon transverse response, focusing exclusively on the
interference between the MEC and OB currents. The pure MEC responses
have been previously computed and shown in various studies to be
negligible, allowing them to be safely disregarded. Here, we describe
the interference terms separately for the different MEC components:
Seagull, pionic, and \(\Delta\), in combination with the magnetization
and convection terms of the OB currents. This separation is essential
to analyze the relative contributions of each component to the overall
response.  

It should be clarified that here we are
computing the single-nucleon response corresponding to either a proton
or a neutron, with the requirement that the isospin of \( p \) and \(
h \) must be the same $t_p=t_h$. At the end of the calculation, the
contributions from protons and neutrons must be summed to obtain the
total response.

The magnetization-seagull (ms) interference is given by Eq. (\ref{w1b2b})
\begin{equation}
w^T_{ms}=w^{11}_{ms}+w^{22}_{ms}= {\rm Re}\sum
\nj_m(p,h)^*\cdot\nj_s(p,h)
\end{equation}
where we use that the magnetization current
is perpendicular to $\nq$ and it has only $x,y$ components. 
Using Eq. (\ref{magnetization}) for the magnetization current we can write
\begin{eqnarray}
w^T_{ms}
&=&
\delta_{t_pt_h}{\rm Re} \sum_{s_ps_h}
(-\frac{G_M^h}{2m_N}i\nq\times\nsigma_{s_ps_h})^*
\cdot \nj_s(p,h)
\nonumber\\
&=&
\sum_{s_ps_h}
\frac{G_M^h}{2m_N}i(\nq\times\nsigma_{s_hs_p})
\cdot \nj_s(\np,\nh)_{s_ps_h}.
\label{mssum}
\end{eqnarray}
We have utilized the fact that the spin sum 
already yields a real number, as will be shown later, making it unnecessary to
explicitly take the real part.

By following the same procedure, we express the various interferences
required between the convection and magnetization currents with the
seagull, pionic, and \(\Delta\) operators, as follows:
\begin{widetext}
\begin{eqnarray}
w^T_{cs} & = & {\rm Re}\sum \nj^T_c(p,h)^*\cdot\nj_s(p,h) =
\sum_{s_ps_h}
\frac{G_E^h}{m_N} \delta_{s_hs_p} \nh_T \cdot \nj_s(\np,\nh)_{s_ps_h}
\label{cssum}
\\
w^T_{m\pi}&  = & {\rm Re}\sum \nj_m(p,h)^*\cdot\nj_\pi(p,h) =
\sum_{s_ps_h}
\frac{G_M^h}{2m_N}i(\nq\times\nsigma_{s_hs_p})
\cdot \nj_\pi(\np,\nh)_{s_ps_h},
\label{mpisum}
\\
w^T_{c\pi} & = & {\rm Re}\sum \nj^T_c(p,h)^*\cdot\nj_\pi(p,h) =
\sum_{s_ps_h}
\frac{G_E^h}{m_N} \delta_{s_hs_p} \nh_T \cdot \nj_\pi(\np,\nh)_{s_ps_h}
\label{cpisum}
\\
w^T_{m\Delta}&  = & {\rm Re}\sum \nj_m(p,h)^*\cdot\nj_\Delta(p,h) =
\sum_{s_ps_h}
\frac{G_M^h}{2m_N}i(\nq\times\nsigma_{s_hs_p})
\cdot \nj_\Delta(\np,\nh)_{s_ps_h}
\label{mdsum}
\\
w^T_{c\Delta}&  = & 0.
\end{eqnarray}
Note that only the transverse component of the convection current
appears that is proportional to the transverse nucleon momentum,
$\nh_T=\nh-\frac{\nh\cdot\nq}{q^2}\nq$,
thereby selecting the \(x\) and \(y\) components in the
transverse response. We also anticipate that the convection-\(\Delta\)
interference is zero because the convection current is
spin-independent, while the \(\Delta\) current is linear in the
\(\sigma\) operators.

The sums over spin in Eqs. (\ref{mssum}--\ref{mdsum}) 
are performed in Appendix \ref{AppendixD}. 
\begin{eqnarray}
w^T_{ms}(p,h)
&=&
4t_h\frac{f^2}{m_\pi^2}F_1^V\frac{G_M^h}{2m_N}
\int \frac{d^3k}{(2\pi)^3}
\left(
\frac{4\nq\cdot\nk_1}{\nk_1^2+m_{\pi}^2}+
\frac{4\nq\cdot\nk_2}{\nk_2^2+m_{\pi}^2}
\right)
\equiv
4t_h\frac{f^2}{m_\pi^2}F_1^V\frac{G_M^h}{2m_N}\  {\cal I}_{ms}(\np,\nh)
\label{wms}
\\
w^T_{cs}(p,h)
&=&
4t_h\frac{f^2}{m_\pi^2}F_1^V\frac{G_E^h}{m_N}
\int \frac{d^3k}{(2\pi)^3}
\left(
\frac{2\nh_T\cdot\nk_1}{\nk_1^2+m_{\pi}^2}-
\frac{2\nh_T\cdot\nk_2}{\nk_2^2+m_{\pi}^2}
\right)
\equiv
4t_h\frac{f^2}{m_\pi^2}F_1^V\frac{G_E^h}{m_N}\ {\cal I}_{cs}(\np,\nh)
\\
w^T_{m\pi}(p,h)
&=&
-4t_h\frac{f^2}{m_\pi^2}F_1^V\frac{G_M^h}{2m_N}
\int \frac{d^3k}{(2\pi)^3}
\frac{4(\nq\times\nk_2)^2}{(\nk_1^2+m_{\pi}^2)(\nk_2^2+m_{\pi}^2)}
\equiv
-4t_h\frac{f^2}{m_\pi^2}F_1^V\frac{G_M^h}{2m_N}\  {\cal I}_{m\pi}(\np,\nh)
\label{mpi}
\\
w^T_{c\pi}(p,h)
&=&
-4t_h\frac{f^2}{m_\pi^2}F_1^V\frac{G_E^h}{m_N}
\int \frac{d^3k}{(2\pi)^3}
\frac{4(\nq\cdot\nk_2-\nk_2^2)\nh_T\cdot\nk_2}{(\nk_1^2+m_{\pi}^2)(\nk_2^2+m_{\pi}^2)}
\equiv
-4t_h\frac{f^2}{m_\pi^2}F_1^V\frac{G_E^h}{m_N}\
 {\cal I}_{c\pi}(\np,\nh)
\\
w^T_{m\Delta}(p,h)
&=&
-4t_h C_\Delta \frac{G_M^h}{2m_N}
\int \frac{d^3k}{(2\pi)^3}
2\left(
\frac{3q^2k_1^2-(\nq\cdot\nk_1)^2}{\nk_1^2+m_{\pi}^2}+
\frac{3q^2k_2^2-(\nq\cdot\nk_2)^2}{\nk_2^2+m_{\pi}^2}
\right)
\equiv
-4t_h C_\Delta \frac{G_M^h}{2m_N}\
 {\cal I}_{m\Delta}(\np,\nh),
\nonumber\\
\label{wmd}
\end{eqnarray}
where $C_\Delta$ is defined in Eq. (\ref{cdelta}), $\nk_1=\np-\nk$.
and $\nk_2=\nk-\nh$.  In Eqs. (\ref{wms}--\ref{wmd}) we have defined
the integrals ${\cal I}_{ab}(\np,\nh)$, that are spin independent.

Finally, the total interference between the one-body and two-body
currents is given by the sum of the individual interferences between
the different terms of the currents, namely the seagull, pionic, and
Delta contributions with magnetization and convection currents.
\begin{equation}
w^T_{1b2b}=
w^T_{ms}+
w^T_{cs}+
w^T_{m\pi}+
w^T_{c\pi}+
w^T_{m\Delta}.
\end{equation}

\end{widetext}

\subsection{Low-momentum results}

\newtheorem{theorem}{Result}

\begin{theorem}
Let $\mathbf{q}$ and $\omega$ denote the momentum and energy transfer
in a quasielastic lepton-nucleus scattering process. In the
non-relativistic Fermi gas model, and using non-relativistic
expressions for both the one-body (OB) current and the $\Delta$-isobar
meson-exchange current, the transverse interference response function
$w^T_{m\Delta}(\mathbf{q}, \omega)$ satisfies
\begin{equation}
w^T_{m\Delta}(\mathbf{q}, \omega) < 0
\end{equation}
That is, under moderate momentum and energy transfer
conditions, the interference between the OB and $\Delta$ currents gives a
negative contribution to the transverse response.
\end{theorem}

To demonstrate the result, we first need to express the total
single-nucleon interference responses as the sum of the contributions
from protons and neutrons.
\begin{eqnarray}
w^T_{ms}(\np,\nh)
&=&
\frac{f^2}{m_\pi^2}F_1^V\frac{G_M^p-G_M^n}{m_N}\  {\cal I}_{ms}(\np,\nh)
\label{wmstot}
\\
w^T_{cs}(\np,\nh)
&=&
2\frac{f^2}{m_\pi^2}F_1^V\frac{G_E^p-G_E^n}{m_N}\ {\cal I}_{cs}(\np,\nh)
\\
w^T_{m\pi}(\np,\nh)
&=&
-\frac{f^2}{m_\pi^2}F_1^V\frac{G_M^p-G_M^n}{m_N}\  {\cal I}_{m\pi}(\np,\nh)
\label{wmpitot}
\\
w^T_{c\pi}(\np,\nh)
&=&
-2\frac{f^2}{m_\pi^2}F_1^V\frac{G_E^p-G_E^n}{m_N}\
 {\cal I}_{c\pi}(\np,\nh)
\\
w^T_{m\Delta}(\np,\nh)
&=&
-  \sqrt{ \frac32 } \frac29 
\frac{ff^*}{m_\pi^2}
\frac{C_3^V}{m_N^2}\frac{G_M^p-G_M^n}{m_\Delta-m_N}
 {\cal I}_{m\Delta}(\np,\nh).
\nonumber\\
\label{wmdtot}
\end{eqnarray}

It suffices to verify that the
single-nucleon interference response \( w^{T}_{m\Delta} < 0 \)
in Eq (\ref{wmdtot}). 
On the one hand, \( w^T_{m\Delta} \) is proportional to
the integral \( {\cal I}_{m\Delta}\), 
which contains the pion propagator multiplied by a factor 
that is
always positive.  
The term in question can be seen inside the integral of Eq. (\ref{wmd}),
 given by 
\begin{equation}
3q^2k_i^2-(\nq\cdot\nk_i)^2 \geq 0,
\end{equation}
with $i=1,2$.
This ensures that the integral \( {\cal I}_{m\Delta} \geq 0\). On the other
hand, \( w^T_{m\Delta} \) is also proportional to \( G_M^p - G_M^n \),
which is positive as well.  Therefore, since \( w^T_{m\Delta} \) includes an
overall negative sign, the final result is negative, completing the
proof.

Typically, the result is valid for momentum transfers below
approximately 500 MeV, where the interference response $m\Delta$ is
explicitly negative.  For momentum transfers above this threshold,
relativistic effects become increasingly significant. In this regime,
the explicit determination of the sign is no longer straightforward
due to the complexity of the spin sum in the relativistic case. The
analysis requires numerical computations to verify the sign of the
interference, as the non-relativistic result no longer applies
directly.

From Eq. (\ref{mpi}), we can also establish the following result for
the magnetization-pionic response:

\begin{theorem}
The transverse interference response between the pionic current and
the magnetization current is negative in the non-relativistic Fermi
gas: $w^T_{m\pi}<0$

\end{theorem}

The proof of this result follows similarly to Result 1, by noticing
that the integral \(\mathcal{I}_{m\pi}\) is positive, as it contains
the square of \(\mathbf{q} \times \mathbf{k}_2\), as seen in
Eq. (\ref{mpi}). Then, according to Eq. (\ref{wmpitot}), we conclude that \(
w^T_{m\pi} < 0 \).

The convection-pionic interference does not generally have a
well-defined sign, but its contribution is much smaller than the
magnetization interference. Therefore, Result 2 can be approximately
extended to the total pionic-OB interference.

For the seagull-magnetization interference, a rigorous result is also
difficult to establish. However, certain particular cases suggest a
trend. For instance, in the case \( \nh = 0 \), it can be demonstrated
that \( w^T_{ms} > 0 \). Additionally, by analyzing the integrand of
\( \mathcal{I}_{ms} \) for \( \mathbf{k} = 0 \), we observe that it
remains positive below the quasielastic peak and changes sign for \(
\omega > (q^2/2m_N) (1+(2m_\pi/q)^2)^{1/2} \). This suggests a general
tendency: the interference starts positive at small \(\omega\) and
eventually changes sign at some point beyond the quasielastic peak,
though the precise location cannot be determined analytically.

The integrands and signs in the equations for the \( ms \), \( m\pi
\), and \( m\Delta \) transverse responses are consistent with those
in the pioneering work of Kohno and Ohtsuka \cite{Koh81}, which was
among the first to compute 1b2b interferences using Riska's
currents. Similar expressions were also obtained in
\cite{Ama94a,Ama94b}, although written in a different form. One of the
key contributions and novelties of the present work is the observation
that the signs of the \( m\Delta \) and \( m\pi \) contributions are
evidently negative, which follows trivially from Eqs. (98) and (100),
as established in our results.

\section{Results}

Here we present results for the transverse response functions due to
the interference between the MEC and OB current in the 1p1h
channel. As discussed in the previous sections, these interferences are
expressed as an integral of an effective single-nucleon
interference. In the non-relativistic Fermi gas, which we have examined in
great detail, the single-nucleon interferences are represented through
relatively complex integrals after analytically computing the spin
traces. In the case of the seagull and $\Delta$ currents, these integrals
are analytical. For the $\Delta$ current, it has been proven that the
associated response is always negative for all values of \(q\) and
\(\omega\)  (Result 1). In this results section, we calculate the interference transverse
responses for various values of \(q = 300, 400, 550 \, \text{MeV/c}\),
and show the results as a function of \(\omega\) for
\(^{12}\text{C}\). 
 We employ several nuclear models to compare the
responses, primarily aiming to observe if the results deviate or not from
the Fermi gas significantly.
The nuclear models we use include: non-relativistic Fermi gas,
relativistic Fermi gas, mean-field models, semi-relativistic models
(both Fermi gas and mean field), and the spectral function
model. Relativistic mean field and superscaling models
with effective nucleon mass are also considered. The mean-field models
include the Woods-Saxon potential, Dirac-equation-based potential, and
the plane wave approximation. Many of these models have been
previously applied in calculations for the study of electron
scattering. Our results show approximate agreement in both magnitude
and sign of the different MEC contributions. In particular, all
examined models verify the result that the effect of the delta
current is negative for these values of momentum transfer, and the
total MEC effect is small and predominantly negative. 

\begin{figure}[ht]
  \centering
\includegraphics[width=7.5cm,bb=160 380 380 810]{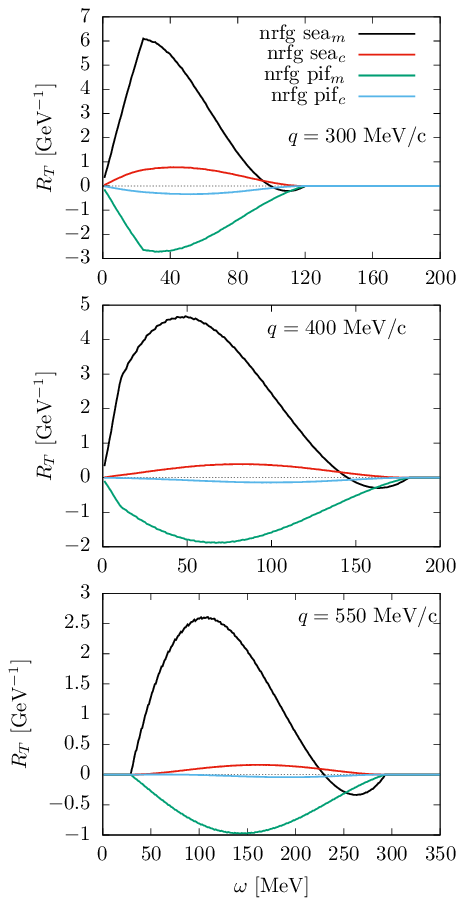}
\caption{ Interferences between the components of the one-body current
  and the MEC in the 1p1h transverse response. Specifically, we represent
  the magnetization-seagull (sea$_m$), convection-seagull (sea{$_c$}),
  magnetization-pionic (pif{$_m$}), and convection-pionic (pif{$_c$}) interferences
  for different values of $q$.}
\label{norel14}
\end{figure}

\subsection{Fermi gas}

We begin by presenting in Fig. \ref{norel14}  results
for the non-relativistic Fermi gas, with $k_F=225$ MeV/c.
 We show the interference of the
seagull and pionic currents with the magnetization and convection
currents. This is done to demonstrate that the contribution of the
convection current in the MEC-OB interference is very small,
particularly in the case of the pionic current. As a result, the
magnetization current is dominant in the interference 
for
these low to intermediate momentum transfer values. Specifically, we
can conclude that, according to Result 2, 
the contribution of the pionic current is negative
if the small convection contribution is disregarded.

Taking this into account, it is worth mentioning the calculation
performed by Alberico et al. \cite{Alb90}.
 In that reference, a positive result was
obtained for the pionic-OB interference, which clearly points to an
error in the calculation, as it also considered a Fermi gas model. As
we have previously discussed, the result establishes that this
interference is negative when convection is neglected. In fact, by
inspecting Eq. (2.41) of Ref. \cite{Alb90}, it can be observed that
the contribution of the pionic current is positive in that reference, 
indicating a possible error in performing the spin summation.

\begin{figure}[ht]
  \centering
\includegraphics[width=7.5cm,bb=160 380 380 810]{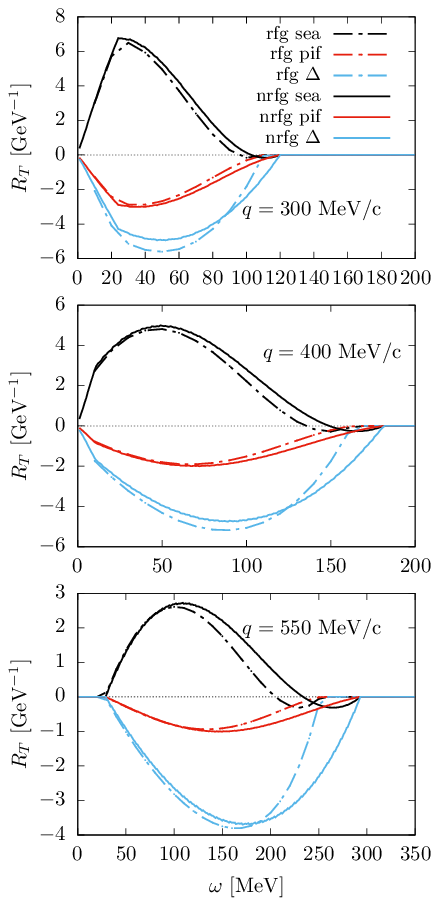}
\caption{ 
Interference between one-body and two-body currents
in the transverse response,
  separated into seagull, pion-in-flight, and $\Delta$ currents.
We  compare the non-relativistic Fermi Gas
  (nrfg) with the Relativistic Fermi Gas (rfg) for three values of the
  momentum transfer in the C12 nucleus. The Fermi momentum is \( k_F =
  225 \) MeV/c.  Diagrams for the 1p1h MEC matrix elements}
\label{norel10}
\end{figure}

In Fig. 4, we present the interferences of the separate MEC
components ---seagull, pionic, and $\Delta$--- with the OB current in the
transverse response. Here, we compare the non-relativistic Fermi gas
to the relativistic Fermi gas. Both models yield similar results, with
increasing differences as the momentum transfer increases, primarily
due to the different kinematics. In fact, it can be checked that the
relativistic result converges numerically to the non-relativistic one
when both the momentum transfer \(q\) and the Fermi momentum approach
zero \cite{Cas23}. Non-relativistic responses extend to higher values
of \(\omega\), due to the kinematics.
We observe that both Results 1 and 2
remain valid in the relativistic case,
even though they wer proven in the non-relativistic limit.

\begin{figure}[ht]
  \centering
\includegraphics[width=7.5cm,bb=160 380 380 810]{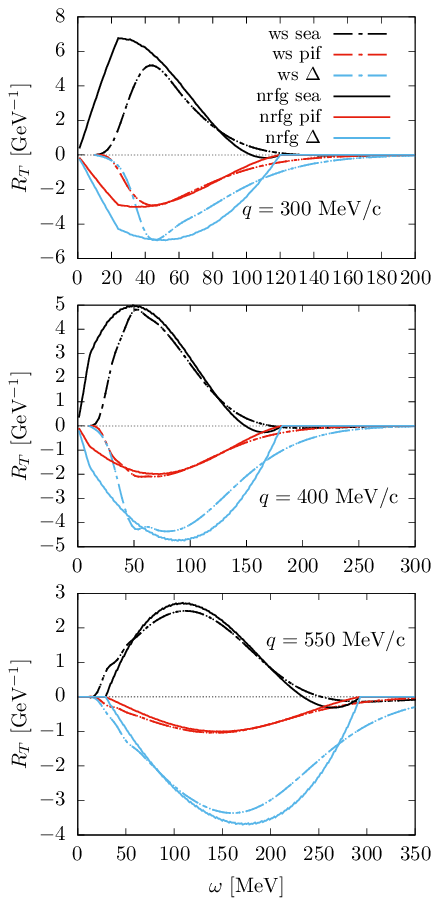}
\caption{ 
The same as Fig. \ref{norel10},
comparing between two models: the mean
  field with Woods-Saxon potential (ws) and the non-relativistic Fermi
  Gas (nrfg) for different values of momentum transfer \( q \) with \(
  k_F = 225 \) MeV/c.
}
\label{norel1}
\end{figure}

\subsection{Mean field with Woods-Saxon potential}

In Fig. 5, we compare the Fermi gas results to those of the mean field
model for finite nuclei, using a Woods-Saxon (WS) potential
\cite{Ama94a,Ama94b}. In this model, the initial and final nucleon
states are solutions to the Schr\"odinger equation
\begin{equation}
\left[ -\frac{1}{2m_N}\nabla^2+V(r)\right]\psi(\nr)=\epsilon \psi(\nr),
\end{equation}
for positive and negative values of the energy $\epsilon$. 
The WS potential is given by 
\begin{equation}
V(r)= -V_0f(r)
+\left(\frac{\hbar}{m_\pi c}\right)^2\frac{\nl\cdot\nsigma}{r}
\frac{df}{dr}+V_C(r),
\end{equation}
where whe function $f(r)$ is the standard Woods-Saxon shape function
\begin{equation}
f(r)= \frac{1}{1+e^{(r-R)/a}},
\end{equation}
and  $V_C(r)$ in the Coulomb potential of a homogeneously charged sphere
with radius $R$. For $^{12}$C 
we use the  WS parameters for $^{12}$C given in Table 1.
\begin{table}
\caption{Parameters of the Woods-Saxon potential used in this work for
  the nucleus $^{12}$C.}
\begin{ruledtabular}
\begin{tabular}{ccccc}
         &$V_0$ [MeV] & $V_{ls}$ [MeV]& $a$ [fm] & $R$ [fm] \\ \hline
protons  &$62$         & $3.2$            & 0.57   &  2.86 \\
neutrons &$60$         & $3.15$            & 0.57   &  2.86 \\
\end{tabular}
\end{ruledtabular}
\end{table}

For $^{12}$C, the initial states include nucleons in the occupied shells
$1s_{1/2}$ and
$1p_{3/2}$. More details can be found in
Refs. \cite{Ama94a,Ama94b}. Note that there is a
typographic error in Ref. \cite{Ama94a} regarding the relative sign
between the central and spin-orbit potentials. This is merely a
mistake in the written formula and does not affect the results. The
energy of the \(1p_{3/2}\) state is lower than that of the
\(1p_{1/2}\) state because the spin-orbit potential is negative for
the \(1p_{3/2}\) state and positive for the \(1p_{1/2}\) state.

The mean-field  approach accounts for some effects of
the final-state interaction (FSI) in the response functions. 
Additionally, unlike the Fermi gas
model, it incorporates finite-size effects along with surface effects
of the nucleus.
In Fig. 5 we observe some differences between the WS model and the Fermi gas,
particularly at low momentum transfer and low energy, where Pauli
blocking affects the Fermi gas more significantly. The WS response
shows a slight tail extending to higher energies, unlike the Fermi
gas. However, at \(q = 550\) MeV/c, the two models become more
similar, except for the high-energy tail seen in the WS model. A
possible explanation for this similarity at intermediate momentum is
that the wavelength of the exchanged photon is small compared to both
the nuclear surface and the extent of the nucleon orbits or 
wave functions in the
occupied shells.

\begin{figure}[ht]
  \centering
\includegraphics[width=7.5cm,bb=160 380 380 810]{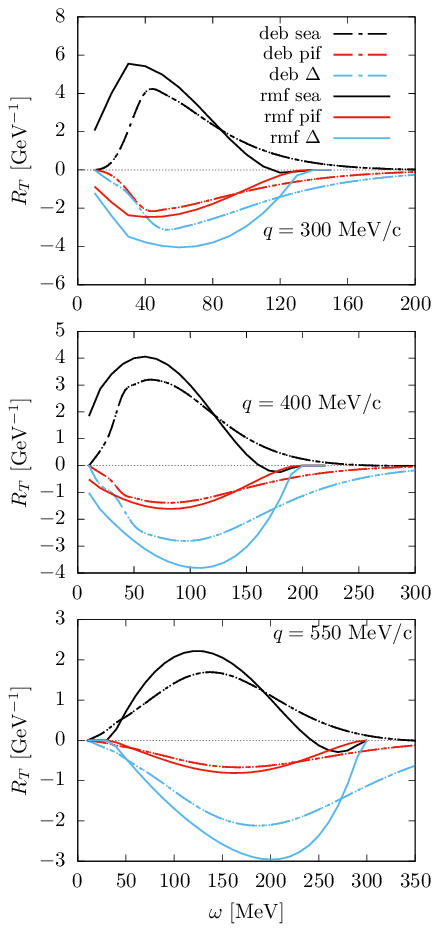}
\caption{ 
The same as Fig. \ref{norel10}.
Results are compared between two models: the
  relativistic mean field with DEB potential (deb) and the
  relativistic mean field of nuclear matter (rmf) 
  with effective mass \( M^* = 0.8 \), for different values of
  momentum transfer \( q \).
}
\label{norel2}
\end{figure}

\subsection{Relativistic mean field}

In fig. \ref{norel2} we present the interference responses
for  two relativistic mean field models:
  the  Dirac-equation based (DEB) model  and the
  relativistic mean field of nuclear matter 
  with effective mass.

In the  RMF model nucleons
move in the presence of scalar \( U_S(\mathbf{r}) \) and vector \(
U_V(\mathbf{r}) \) potentials, and satisfy a Dirac-like equation for
the four-component nucleon wave function \(\Psi(\mathbf{r})\):
\begin{equation} \label{dirac}
\left[\gamma^0\left(E - U_V(\mathbf{r})\right) 
- \boldsymbol{\gamma} \cdot \mathbf{p} 
- \left(M + U_S(\mathbf{r})\right)\right] \Psi(\mathbf{r}) = 0.
\end{equation}
where $\Psi$ has up and down components
\begin{equation}
\Psi(\nr)= \left(
\begin{array}{c}
\psi_{u}(\nr)
\\
\psi_{d}(\nr)
\end{array}
\right).
\end{equation}
The DEB potential is obtained by rewriting the Dirac equation (\ref{dirac})
as a  Klein-Gordon equation for the upper
component of the wave function. In this reduction, the upper component
is written in the form
\begin{equation}
\Psi_{u}(\mathbf{r}) = A^{1/2}(r,E)\, \phi(\mathbf{r}),
\end{equation}
where \(E\) is the energy in the final state and \( A(r,E)\)
is the Darwin term
\begin{equation}
A(r,E)=1+ \frac{U_S(r)-U_V(r)}{E+M}.
\end{equation}
With this definition the function $\phi(\nr)$ verifies 
the equation
\begin{equation}
\left[ -\frac{1}{2m_N}\nabla^2+ U_{DEB}(r,E)\right] \phi(\nr)=
\frac{E^2-m_N^2}{2m_N}\phi(\nr).
\end{equation}
The DEB potential is given by \cite{Hor91,Udi95}
\begin{equation}
V_{DEB}=V_C + V_{so}\nsigma\cdot\nl+V_D+V_{coul}
\end{equation}
where the central, spin-orbit and Darwin potentials are given by
\begin{eqnarray}
V_C(r,E) &=& 2m_NU_S(r)+2EU_V(r)+U_S(r)^2-U_V(r)^2 
\nonumber\\
V_{so}(r,E) &=& -\frac{1}{rA}\frac{\partial A}{\partial r}
\nonumber \\
V_D(r,E) &=& 
         \frac{3}{4A^2}\left(\frac{\partial A}{\partial r}\right)^2
        -\frac{1}{rA}\frac{\partial A}{\partial r}
        -\frac{1}{2A}\frac{\partial^2 A}{\partial r^2},
\nonumber
\end{eqnarray}
and $V_{coul}$ is the Coulomb potential of a homogeneously charged sphere 
with nuclear radius $R$.

In Fig. \ref{norel2} we present the results of the interference 1b2b
transverse response using the DEB potential within the
semi-relativistic model of Refs. \cite{Ama10b,Ama05b}. It is observed
that the contribution from the Delta current is negative, as is the
contribution from the pion-in-flight current. Consequently, this model
verifies the low momentum results.

In the same figure \ref{norel2}, the results using the DEB potential
are compared with those obtained from the RMF of nuclear matter
\cite{Cas23}. In this model, the scalar and vector potentials are
constant, making it similar to the RFG but with the nucleon mass
replaced by an effective mass \(m_N^*=m_N+U_S\) and the energy
increased by a constant vector energy \(E_V=U_V\). For the
\(^{12}\text{C}\) case shown in Fig. 6, the values used are \(m_N^* =
0.8\,m_N\) and \(E_V = 141\) MeV. More details of the RMF nuclear
matter model with MEC can be found in Refs. \cite{Cas23,Mar21}.  As
seen in Fig. 6, both the DEB model and the RMF model with effective
mass yield qualitatively similar results, with the peaks of the
interference responses largely coinciding. This similarity arises
because both models incorporate final-state interaction
effects. However, the absolute values obtained with the DEB model are
smaller. This is a consequence of the fact that, in the DEB model, the
effective mass depends on \(r\), leading to responses that exhibit a
tail extending much further than those in the shell model or nuclear
matter. Essentially, it appears as if the strength is spread over a
wider energy interval.  In any case, it is remarkable that the low
momentum results remain verified in the models presented in Fig. 6:
the 1b-$\Delta$ interference is negative and the 1b-pionic one is negative.

\begin{figure}[ht]
  \centering
\includegraphics[width=7.5cm,bb=180 380 430 810]{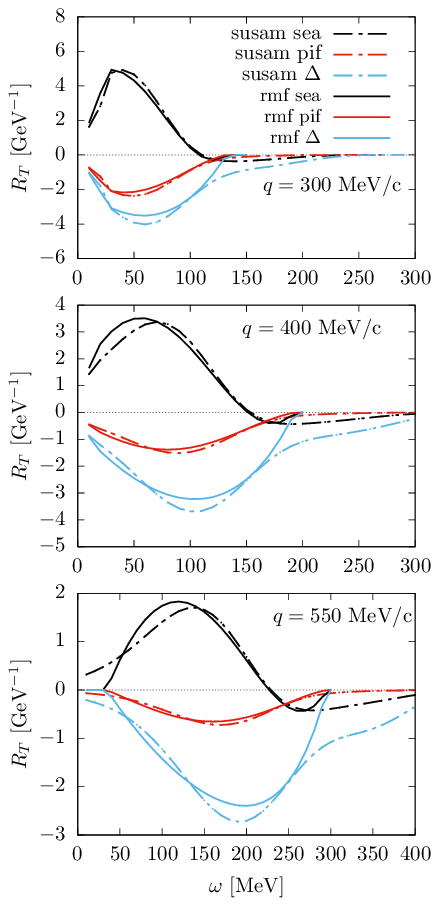}
\caption{ 
The same as Fig. \ref{norel10}.
Results are compared between two models: the
  superscaling model (susam) with relativistic effective mass
and the relativistic mean field of nuclear matter (rmf) 
  with effective mass \( M^* = 0.8 \), for different values of
  momentum transfer \( q \).
}
\label{norel15}
\end{figure}

\subsection{SuSAM* model}
In Fig. 7, we compare the RMF of nuclear matter with the
superscaling model with effective mass (SuSAM*) as described in
Ref. \cite{Cas23}. The SuSAM* model is an extension of the
superscaling (SuSA) model presented in Ref. \cite{Ama05} and employs a
phenomenological scaling function fitted to the
quasielastic cross section data. In this approach, the cross section
is approximated via factorization as the product of the
phenomenological scaling function and a single-nucleon cross section,
which is conveniently derived from the RMF equations
of nuclear matter with effective mass,
allowing for a unified description of the data with a single scaling
function. Recently, the model was further refined to improve the
single-nucleon prefactor by defining it as an average value of the
nucleon response in the Fermi gas. This new definition enables the
extension of the average response into the high-momentum region, beyond
the conventional Fermi momentum, by replacing the Fermi gas momentum
distribution by a smeared distribution effectively including a gradual
rather than abrupt transition of the nucleon momentum distribution at
the Fermi surface \cite{Cas23a}. Instead of a sharp cutoff at the
Fermi momentum, the Fermi
surface is diffused over a range of momenta.

This smeared momentum distribution allows for the definition of a
positive-definite averaged single-nucleon cross section, replacing the
simple extrapolation of the Fermi gas average which loses its meaning
outside the kinematically allowed range of the Fermi gas. Moreover,
this novel procedure enabled the extension of the model in
Ref. \cite{Cas23} to include MEC within the SuSAM* formalism by using
the same smeared momentum distribution. The results of the SuSAM*
model for the 1b-MEC interference are compared in Fig. 7 with those of
the RMF model with effective mass. As shown in the figure, the scaling
model produces responses similar to the RMF; however, the responses in
the scaling model extend beyond the interval permitted by the Fermi
gas. In any case, the model continues to verify the low momentum
results.

\subsection{Strong form factor and relativistic effects}

\begin{figure}[ht]
  \centering
\includegraphics[width=7.5cm,bb=160 380 380 810]{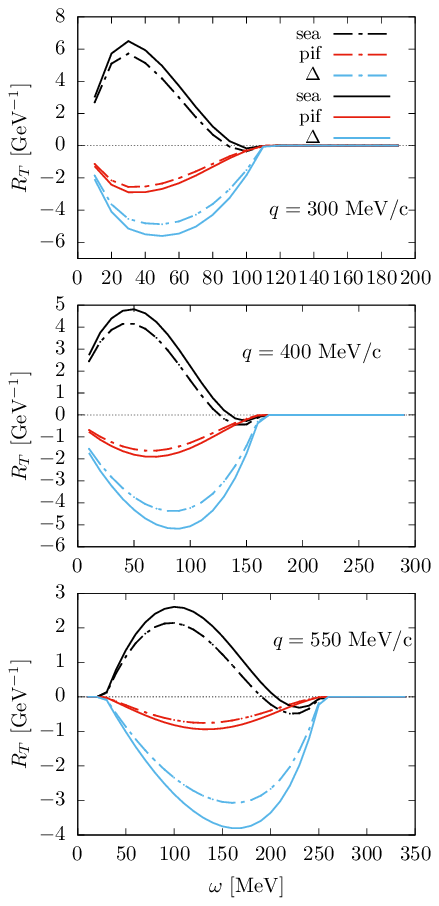}
\caption{  The same as Fig. \ref{norel10}
 but calculated with RFG, with and without the strong $\pi NN$ form factor. The dashed line represents the results with the form factor, 
while the solid line represents the results without it.
}
\label{norel5}
\end{figure}

In Fig. \ref{norel5}, we show the effect of including the \(\pi NN\) and \(\pi
N\Delta\) form factors. In our non-relativistic Fermi gas equations
and in the low-\(q\) results, we have omitted these form factors for
simplicity. These form factors are multiplicative factors that would
be included inside the internal integrals over the intermediate
nucleon momentum \(\mathbf{k}\). Their inclusion does not affect the
low momentum results since these form factors are positive and do not
alter the sign of the interference. Moreover, in the non-relativistic
calculation, some integrals can be evaluated analytically without the
form factors, which further simplifies the computation. Given that we
are considering small momentum transfers, the effect of the form
factors is minimal, as demonstrated in Fig. \ref{norel5}, where the relativistic
Fermi gas results are compared with and without the strong form
factor. As the form factor is less than one, the inclusion produces a
reduction of the maximum in absolute value.

\begin{figure}[ht]
  \centering
\includegraphics[width=7.5cm,bb=160 380 380 810]{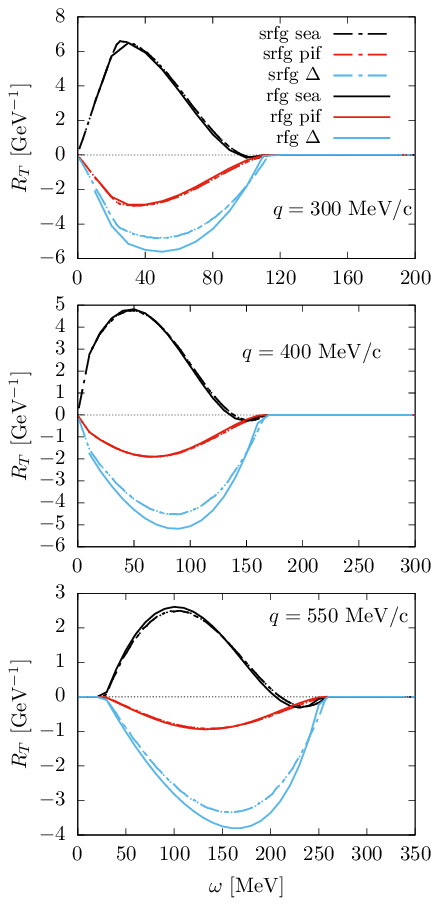}
\caption{ 
The same as Fig. \ref{norel10} but showing two models: relativistic  Fermi gas  (rfg)
  and semirelativistic Fermi gas (srfg). The comparison illustrates
  the differences between the relativistic and semirelativistic
  approaches in the transverse response, for different values of $q$.  }
\label{norel6}
\end{figure}

The relativistic Fermi gas can be compared with the semirelativistic
Fermi gas model (SRFG) developed in \cite{Ama03}. The SRFG model
starts from the non-relativistic Fermi gas, incorporating relativistic
kinematics and replacing the non-relativistic current with a
semirelativistic expansion in powers of the initial nucleon momentum
divided by the nucleon mass $(h/m)$, while preserving the exact
dependence on the final momentum. This approach was extended to
include MEC \cite{Ama02} and also applied to the Delta current,
although in the latter case the semirelativistic correction is not
exact due to the use of a static Delta propagator.  The
semirelativistic current is obtained from the relativistic one by
multiplying by a factor \(1/\sqrt{1+\tau}\).  In Fig. \ref{norel6},
the SRFG model is compared with the exact RFG for the interference
between the MEC and the one-body current. For the seagull and pionic
contributions, the SRFG model agrees very well with the relativistic
one. However, for the Delta contribution, the
approximation is less accurate because the static Delta propagator
limits the effectiveness of the semirelativistic factor.

In addition, the semirelativistic model was extended to be applied in
conjunction with the Woods-Saxon mean field model \cite{Ama10b}. This
extended model can be directly compared with the DEB model. In fact,
the DEB model also incorporates semirelativistic MEC currents, but
these currents are further modified because the pion propagator in the
DEB model is made dynamic by including the pion energy as the
difference in energy between the nuclear states of the mean field
model. The comparison between these two models, DEB and SRWS, as shown
in Fig. 10, reveals significant differences in both the width and the
height of the interference response peak. Specifically, the DEB model
extends to higher energies and exhibits a broader peak, which is
attributed to the fact that the DEB potential is much stronger than
the Woods-Saxon potential.

\begin{figure}[ht]
  \centering
\includegraphics[width=7.5cm,bb=160 380 380 810]{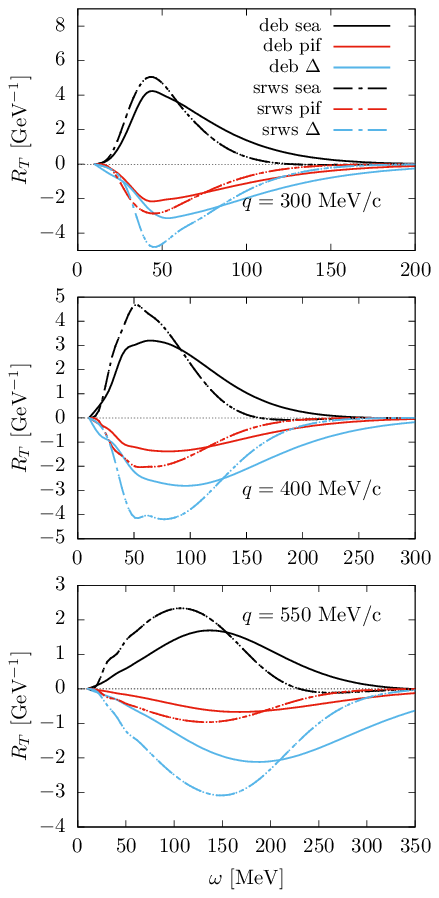}
\caption{ The same as Fig. \ref{norel10} , but now comparing the
  models: relativistic mean field with DEB potential (deb) and
  semi-relativistic mean field with Woods-Saxon potential (srws) for
  different values of momentum transfer \( q \).
}
\label{norel3}
\end{figure}

\subsection{Plane wave approximation}

In the shell model with a Woods-Saxon potential, the plane wave
approximation (PWA) assumes that the final nucleon with momentum \(\np\) is
described by a plane wave, meaning it is a solution of the Schrödinger
equation without final-state interactions. 
Note that in PWA the sum over hole states, \(h\) in Eq. (\ref{hadronic}) 
refers to a sum over
occupied states in the shell model, just as the sum over spectator
states \(k\) in Eq. (\ref{melement}) also corresponds to occupied Woods-Saxon
states. Therefore, in this approach, the plane wave approximation is
applied only to the final outgoing particle state, while the initial
state nucleons remain described by the bound shell model wave
functions.

Results using this model are presented in Fig. 11, where they are
compared with the Woods-Saxon mean-field calculations for the 1b2b
interference responses. The observed effect is similar to that seen in
the 1b response within the Plane Wave Impulse Approximation (PWIA)
\cite{Ama05b}. The impact of final-state interactions appears as a
shift in the response. This shift can be understood as a consequence
of the energy imbalance between the initial and final states. In the
initial state, the nucleon has both kinetic and potential energy,
whereas in the final state, only kinetic energy remains, since the
potential is neglected. This energy mismatch propagates to the
energy-conserving delta function, altering the position of the
response peak.

\begin{figure}[ht]
  \centering
\includegraphics[width=7.5cm,bb=160 380 380 810]{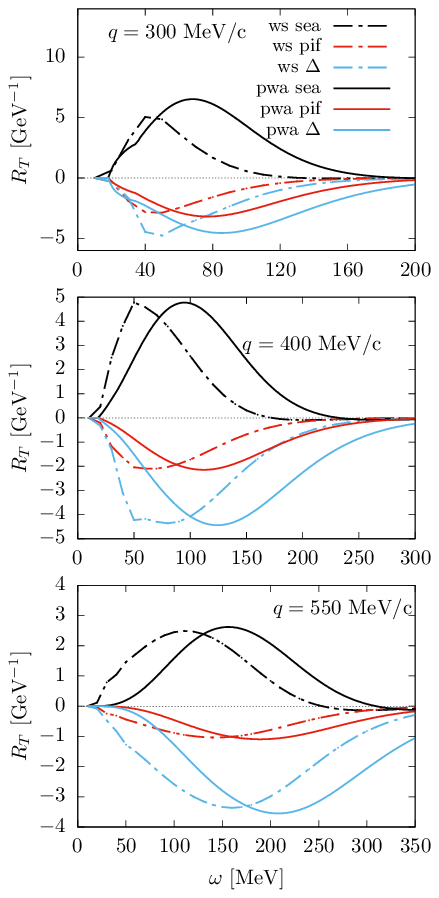}
\caption{ The same as Fig. \ref{norel10}, but now comparing the models: mean
  field with Woods-Saxon potential (ws) and mean field in plane-wave
  approximation for the final state (pwa) for different values of
  momentum transfer \( q \).
}
\label{norel12}
\end{figure}

The shift in the response function can be qualitatively understood
using a back-of-the-envelope estimate. First we assume that the matrix
element of the current in PWA is
approximately equal to the matrix element in the Woods-Saxon model,
$\langle J^\mu \rangle_{PW} \simeq \langle J^\mu \rangle_{WS}$.
Second, we approximate the potential energy of the final-state
nucleon as a constant, $V_p \simeq -V < 0$.  Thus, the total energy of
the outgoing particle can be written as the sum of its kinetic and
potential energy: $\epsilon_p = t_p - V.$ Using this, the transverse response
function in PWA can be expressed as
\begin{eqnarray}
R^T_{PW}(q,\omega) &=& \sum_{ph} |\langle J_T \rangle_{PW}|^2
\delta(t_p - \epsilon_h - \omega) \nonumber
\\ 
&\simeq& \sum_{ph}
|\langle J_T \rangle_{WS}|^2 \delta(\epsilon_p+V- \epsilon_h - \omega)
\nonumber
\\ &=& R^T_{WS}(q,\omega-V).
\end{eqnarray}
This expression shows that the response function is effectively
shifted due to the neglect of the potential in the final state.

From Fig. 11, we observe again that the low-momentum result for the
1b-MEC interference response remains valid in both the plane-wave
approximation and the Woods-Saxon potential.

\subsection{Spectral function model}

In this subsection, we present results using the spectral function (SF)
model, which employs the one-hole spectral function, $S(\np,E)$, that depends on
the missing momentum and missing energy.
In the SF model, the transverse response is computed assuming factorization
of the single-nucleon response and the one-hole spectral function for
one-particle emission. 
\begin{equation} \label{spectral}
R_T(q,\omega)= \int d^3 p \, w_T(\np,\np-\nq) S(\np-\nq,\omega-T_p)
\end{equation}
where the single nucleon response is $w^T=w^{11}+w^{22}$, while
 $w^{\mu\mu}$ is defined in Eq. (\ref{single-nucleon}).

 Unlike the single-particle model that assumes holes with definite
 energy, the SF approach accounts for a continuous distribution of
 hole energies. It provides the probability that the system contains a
 hole state with momentum \(\mathbf{h} =
 \mathbf{p} - \mathbf{q}\) and a missing energy \(E = \omega - T_p\),
 where \(T_p = \mathbf{p}^2/(2m_N)\).  The basic theory of the SF
 approach to QE electron scattering is summarized in Appendix \ref{AppendixE}.

\begin{figure}[ht]
  \centering
\includegraphics[angle=-90, width=8.5cm,bb=145 80 520 620]{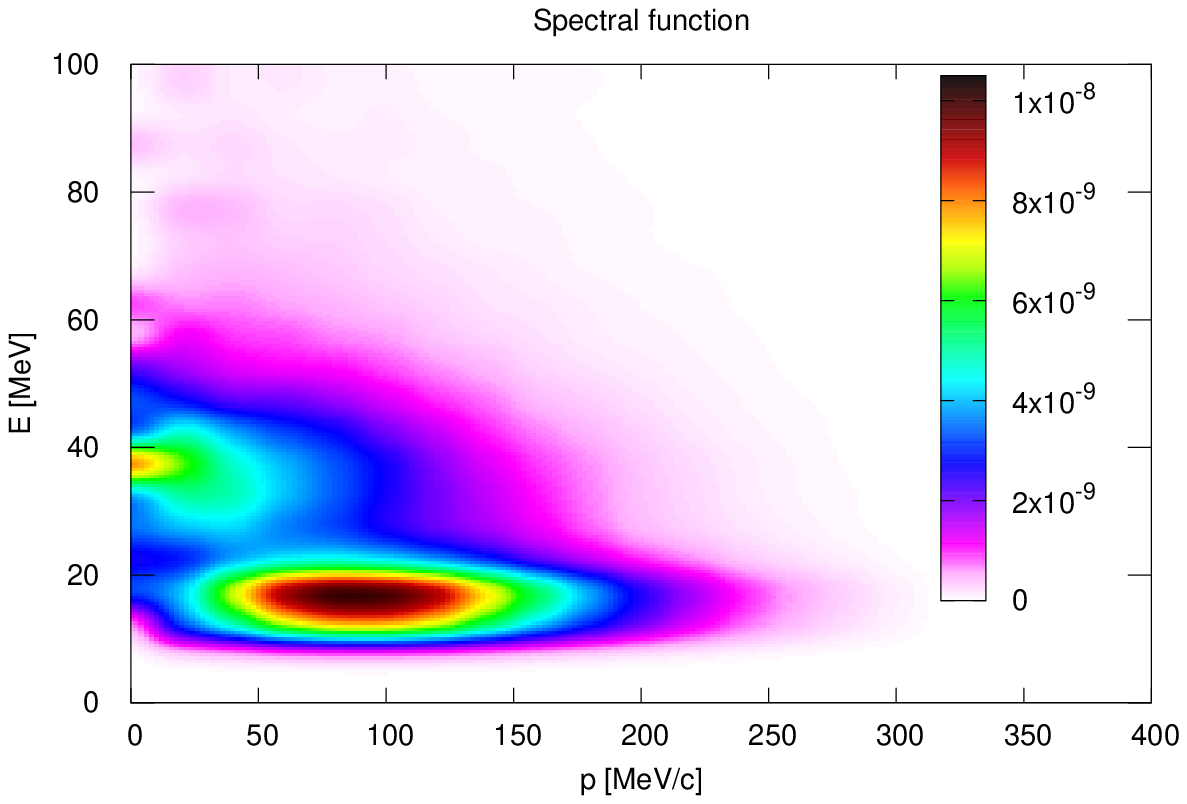}
\caption{Spectral function of $^{12}$C in units of MeV$^{-4} c^3$
}
\label{sf1c}
\end{figure}

We use the spectral function, $S(p,E)$, for \(^{12}\)C taken from
Ref. \cite{valerio} for both protons and neutrons, as shown in
Fig. \ref{sf1c}. This spectral function exhibits peaks around \(E
\simeq 19\) MeV and \(E \simeq 39\) MeV as a function of energy. These
values are close to the binding energies of the \(1p_{3/2}\) and
\(1s_{1/2}\) shells in the extreme shell model, where the nuclear
wave function is described by a Slater determinant.

In the shell model, the spectral function is given by  
\begin{equation}
S(p,E) = \sum_{nlj} (2j+1) |\tilde{R}_{nlj}(p)|^2 \delta(E + \epsilon_{nlj})
\end{equation}
where the sum runs over occupied shells, and \( \tilde{R}_{nlj}(h) \) are the
shell radial wave functions in momentum space, with single-particle energy
$\epsilon_{nlj}$. In the more realistic
spectral function of Fig. 12, the energy dependence is smeared around
the shell binding energies, resulting in a continuous energy
distribution instead of discrete shell levels.

\begin{figure}[ht]
  \centering
\includegraphics[width=7.5cm,bb=100 180 410 770]{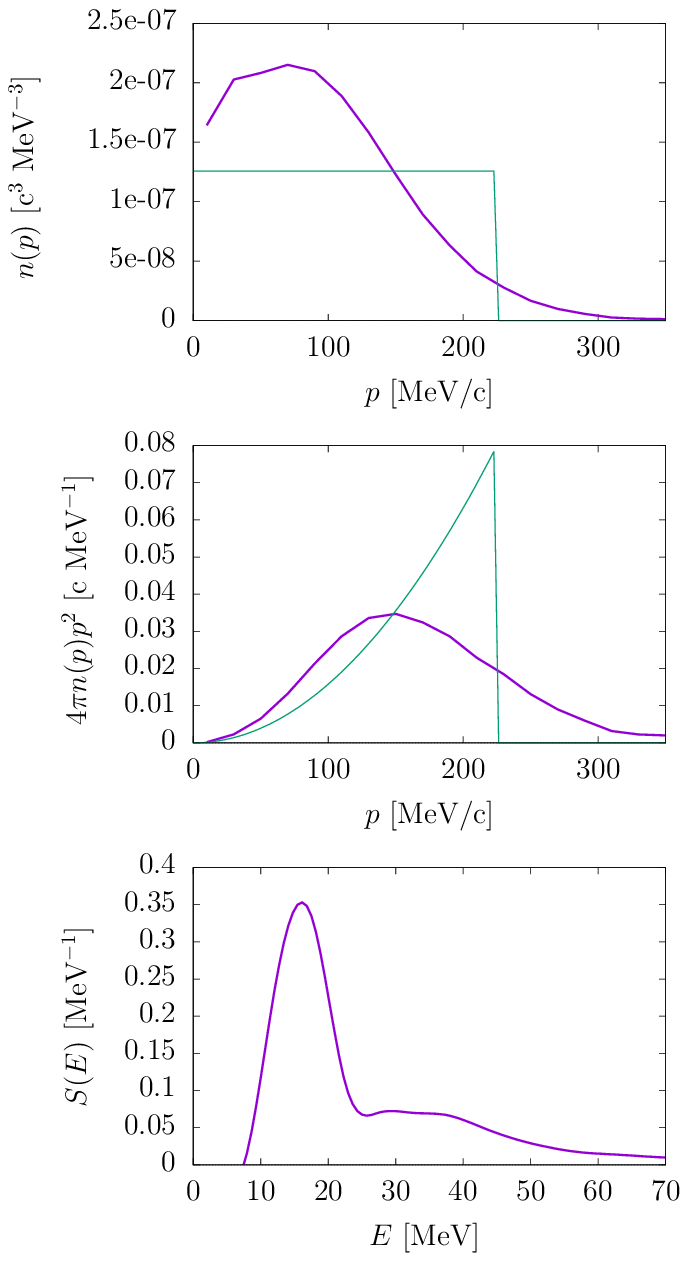}
\caption{ Proton momentum distribution of $^{12}$C (top),  radial
  momentum distribution  (middle) and missing energy distribution (bottom), obtained
  from the spectral function by integration.  }
\label{fig6}
\end{figure}

In Fig. \ref{fig6}, we show the proton momentum distribution $n(p)$ 
obtained by
integrating the spectral function over the missing energy. This
distribution is compared with the constant momentum distribution of
the Fermi gas model. Additionally, we present the radial momentum
distribution, \( 4\pi n(p) p^2 \), which highlights the probability
density of nucleons as a function of momentum. The missing energy
distribution, obtained by integrating the spectral function over
momentum, is also displayed. The normalization follows \( \int d^3p \,
n(p) = 6 \) for \( ^{12}C \), reflecting that the proton and neutron
distributions are identical in this model.

It is worth noting that the response function in the SF model,
Eq. (\ref{spectral}), is expressed as an integral over the final
nucleon momentum \( p \). To evaluate this integral, it is convenient
to first integrate over the missing energy and missing momentum. The
missing energy is given by
\begin{equation}
E = \omega - T_p = \omega - \frac{p^2}{2m_N}.
\end{equation}
Differentiating, we obtain \( dE = -p \, dp / m_N \), and the volume element in spherical coordinates is  
\[
d^3p = m_N \, p \, dE \, d\Omega
\]
where \( \theta \) and \( \phi \) are the nucleon emission angles. The response function can then be rewritten as  
\begin{equation}
R_T(q, \omega) 
= m_N \int dE d\Omega \, p\,   
w_T(\np, \np - \nq) S(\np - \nq, E). 
\label{integral121}
\end{equation}
Next, we define \(\nh = \np - \nq\), leading to the relation  
$h^2 = p^2 + q^2 - 2pq\cos\theta$,
where \(\theta\) is the angle between \(\np\) and \(\nq\), with \(\nq\) chosen along the \(z\)-axis. Differentiating with respect to \(\theta\), we obtain  
$h \, dh = -pq \, d\cos\theta$.

Substituting this into the integral (\ref{integral121}), we can
express the transverse response as
\begin{equation}
R_T(q,\omega) = 2\pi \frac{m}{q} \int_0^\omega dE \int_{|p-q|}^{p+q} S(h,E) w_T(\nh+\nq, \nh) dh,
\end{equation}
where $p=\sqrt{2m_N(\omega-E)}$. Note that $E<\omega$ ensures that $p$
is well defined.  The factor \(2\pi\) arises from the integration over
\(\phi\), and the integration limits in \(h\) correspond to nucleon
emission in the direction of \(\pm \nq\).

The effect of MEC is estimated by treating the spectator nucleon in Eq
(\ref{melement}) as an on-shell plane-wave with momentum $\nk$,
therefore we replace the single-nucleon response by the effective
single nucleon including MEC, in Eq. (\ref{single-nucleon}).  This
approximation has been done in the past in previous calculations by
the Pavia group for $(e,e'p)$ reactions \cite{Bof90}, and in recent
RMF-based calculations \cite{Fra23,Fra25}, where the spectator
nucleon is described using an effective mass and vector energy. A
similar approach to MEC was also adopted in the spectral function
model of Ref. \cite{Lov23}.  Thus the transverse response is evaluated
using the effective single nucleon, Eq. (16), which includes the MEC
contribution, effectively decoupling it from the spectral function.

\begin{figure}[ht]
  \centering
\includegraphics[width=7.5cm,bb=160 380 380 810]{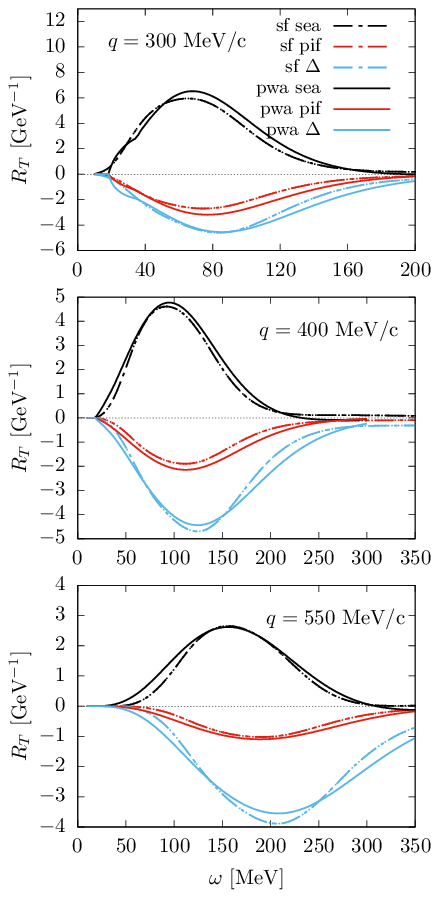}
\caption{ The same as Fig. \ref{norel10}, but now comparing the
  models: spectral function (sf) and mean field with plane-wave
  approximation (pwa) for different values of momentum transfer \( q
  \).}
\label{sf}
\end{figure}

In the spectral function model, the interference MEC-OB responses are
presented in Fig. \ref{sf} for the separate contributions from the seagull,
pion-in-flight, and \(\Delta\) currents. The figure compares the SF
results with those obtained using the PW model from the previous
subsection. Both models yield quite similar results. This similarity
arises from the fact that both models assume plane waves for the
final-state nucleon. In the PW model, the response is obtained by
summing the contributions from each shell separately, while in the SF
model, the shell contributions are smeared according to the spectral
function's energy distribution. However, this smearing effect is
barely noticeable in the inclusive response, as the information about
the hole energy is lost.
Furthermore, the agreement between the SF and PWIA models reinforces
the validity of the approximation that treats the spectator nucleon as
a plane wave. While this approximation is not explicitly made in the
PW model, it is assumed in the SF model. In conclusion, the SF
model, as applied here, fully adheres to the low-momentum result,
consistent with all the models analyzed in this work.

\begin{figure}[ht]
  \centering
\includegraphics[width=7.5cm,bb=180 380 430 810]{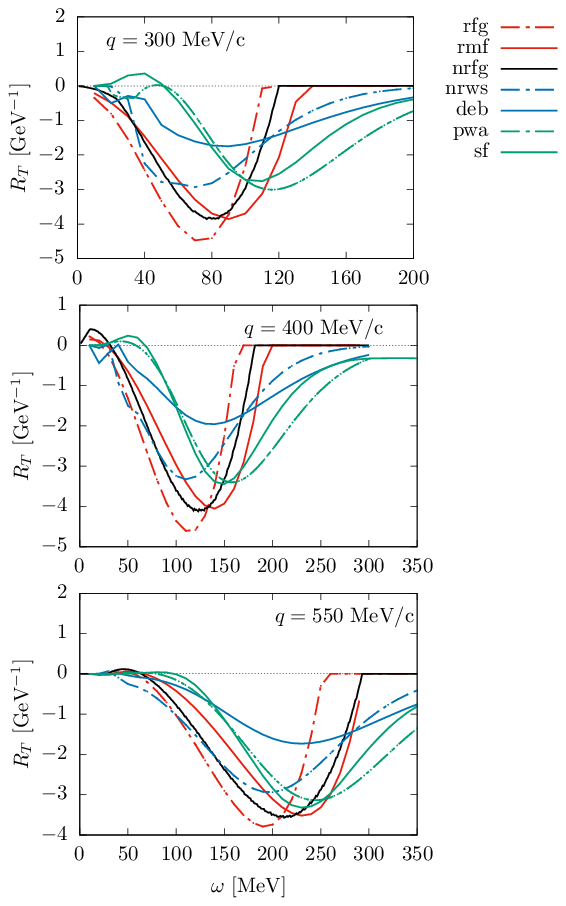}
\caption{
 Total interference OB-MEC compared across all different models considered in this work for all values of $q$. The $\pi NN$ form factor is not included.}
\label{norel13}

\end{figure}

\subsection{Total interference response}

To conclude the results section, we present a comprehensive comparison
of most of the models discussed in this paper to assess the overall
impact of MEC and the theoretical uncertainties. In
Fig. \ref{norel13}, we display the total 1b-MEC interference for a
selection of seven models. All models consistently predict a negative
interference, although there are significant quantitative differences
in the position of the peak and the width of the distribution. Despite
these variations, the overall magnitude remains comparable, with
differences up to a factor of two.  Importantly, the key takeaway from
this comparison ---and one of the main objectives of this work--- is that
the low-momentum result holds across all models analyzed. This
explains why none of the models exhibit a qualitative deviation, such
as a sign change in the interference term.

To end we list here the models considered in this paper:

\begin{enumerate}

\item Non relativistic Fermi gas (nrfg)

\item Relativistic Fermi gas  (rfg)

\item Mean field with Woods-Saxon potential (ws)

\item Mean field with Dirac-equation based potential (deb)

\item Relativistic mean field of nuclear matter with effective mass (rmf)

\item Mean field with plane wave approximation (pwa)

\item Semirelativistic mean field with Woods-Saxon (srws)

\item Semirelativistic Fermi gas (srfg)

\item Superscaling model with effective mass (susam)

\item Spectral function (sf) .

\end{enumerate}

\section{Conclusions}

In this work, we have conducted a detailed reexamination of the OB-MEC
interference in the one-particle emission transverse response,
focusing on its sign. We systematically compared
various models, obtaining qualitatively consistent results. A key
aspect of our analysis was the derivation of two low-momentum results
within the non-relativistic Fermi gas framework, ensuring full
transparency and reproducibility in our approach. The results clearly
establish that the sign of the interference of the one-body current with the pionic and
$\Delta$ currents is negative in the Fermi gas model.

Our results show  also that all models considered satisfy these low-momentum
results. The common feature among these models is that they are based
on independent-particle descriptions ---either relativistic or
non-relativistic--- or extensions, such as the one-hole spectral
function for one-particle emission in electron scattering.

Given these results, it does not seem possible to explain any
enhancement in the transverse response in one-particle emission using
models that do not include fundamentally different ingredients that
would violate the low-momentum results, in contrast with the results
of Refs. \cite{Fra23,Fra25,Lov23}. 
A careful examination of the $\Delta$ current in Refs. [45,46] suggests
  that there is a sign difference with respect to our expression, as
  illustrated in Appendix G.

 A candidate for producing an
enhancement is the inclusion of tensor correlations, as suggested by
the microscopic calculation of Ref. \cite{Fab97}. These correlations
could introduce contributions that go beyond the low-momentum result,
thus altering the single-particle dynamics of the models analyzed
here. However, no study after that of Ref.  \cite{Fab97} has been
conducted to confirm this effect. Future research along these lines is
planned, aiming to include the effect of short-range correlations and
high-momentum components in a two-body independent-pair approximation,
solving the Bethe-Goldstone equation, as outlined in
Ref. \cite{Cas23c}.  In parallel, similar studies for neutrino
scattering are also in progress.

\section{Acknowledgments}

We thank Luis Alvarez-Ruso for helpful discussions.
\vskip 0.2cm

The work was supported by Grant No. 
   PID2023-147072NB-I00
funded by MICIU/AEI /10.13039/501100011033 and by ERDF/EU;
by Grant No. 
   FQM-225 
funded by 
   Junta de Andalucia;
by Grant    NUCSYS
funded by 
    INFN;
by Grant No. 
    BARM-RILO-24-01
funded by 
    University of Turin;
    by  the ``Planes Complementarios de I+D+i" program (Grant ASFAE/2022/022) 
by 
MICIU with funding from the European Union NextGenerationEU and 
Generalitat Valenciana.

\clearpage

\appendix

\section{Lagrangian} 

\label{Appendix0}
In this appendix we display the Lagrangian terms needed in order to obtain the meson-exchange currents. The $\pi NN$ interaction is described using the following Lagrangian

\begin{equation}
\mathcal{L}_{\pi NN}=-\frac{f}{m_\pi}\bar{\Psi}\gamma^\mu\gamma_5  \bm \tau \cdot \partial_\mu\bm \phi \Psi \,,
\end{equation}
where  $\Psi$ represents the nucleon isospinor, and $\bm \phi$ is the isospin triplet pion field $\bm \phi=(\phi_1,\phi_2,\phi_3)$:

\begin{eqnarray}
&&\Psi=\left(\begin{matrix} \psi_p \\ \psi_n \end{matrix}\right)\,,\qquad {\bm \tau \cdot \bm \phi}=\frac{\tau_+\phi + \tau_-\phi^\dagger}{\sqrt{2}} +\tau_3 \phi_3\,, \nonumber \\
&&\tau_\pm\equiv\tau_1\pm i\tau_2\,, \qquad \phi\equiv\frac{\phi_1-i\phi_2}{\sqrt{2}}\,.
\end{eqnarray}

The electromagnetic interactions with the photon field $A_\mu$ are defined as
\begin{equation}
\mathcal{L}_\mathrm{\gamma \pi \pi} =- ieA_\mu \left( \phi^\dagger \partial^\mu \phi - \phi \partial^\mu \phi^\dagger \right)\,,
\end{equation}
\begin{equation}
\mathcal{L}_{\gamma NN\pi} =- ieA_\mu \frac{f}{ m_\pi} \bar{\Psi} \gamma^\mu \gamma_5 \frac{ \tau_+ \phi - \tau_- \phi^\dagger}{\sqrt{2}}  \Psi\,,
\end{equation}
 with $e>0$ the elementary electric charge.
The $\Delta$ interactions are described using the following Lagrangians
\begin{equation}
\mathcal{L}_{\pi N \Delta} = \frac{f^*}{m_\pi} \bar \psi_\mu  \partial^\mu  {\bm \phi} \cdot {\bm T}^\dagger \Psi + h.c.\,,
\end{equation}

\begin{equation}
\mathcal{L}_{\gamma N \Delta}=ie\frac{G}{2m_N}\sqrt{\frac{3}{2}}\bar \psi_\mu \gamma_\nu \gamma_5 T_3^\dagger \Psi F^{\mu \nu} +h.c.
\label{gammaND}\,,
\end{equation}
where $\psi_\mu$ is the $3/2$ spin Rarita-Schwinger $\Delta$ field, $G=2C_{3}^V(Q^2=0)$, with $C_3^V$ defined in Eq.~\eqref{C3}, and $F^{\mu \nu}=\partial^\mu A^\nu-\partial^\nu A^\mu$ the standard electromagnetic tensor.
The $3/2\rightarrow1/2$ isospin transition operator $\bm T$ definition is here reported:
\begin{eqnarray}
T_1 =& \frac{1}{\sqrt{6}} \begin{pmatrix} -\sqrt{3} & 0 & 1 & 0 \\ 0 & -1 & 0 & \sqrt{3} \end{pmatrix} \nonumber\\
T_2 =& -\frac{i}{\sqrt{6}} \begin{pmatrix} \sqrt{3} & 0 & 1 & 0 \\ 0 & 1 & 0 & \sqrt{3} \end{pmatrix} \nonumber\\
T_3 =& \sqrt{\frac{2}{3}} \begin{pmatrix} 0 & 1 & 0 & 0 \\ 0 & 0 & 1 & 0 \end{pmatrix} \,.
\end{eqnarray}
\section{Non relativistic reduction of the $\Delta$ current} 

\label{AppendixA}

Here we perform the non relativistic reduction of the four-vectors
$A^\mu$, and $B^\mu$, Eqs. (\ref{amu},\ref{bmu}), appearing in the
$\Delta$-current. Using the definition of the $\gamma N \Delta$
vertex, Eq. (\ref{gammabetamu}) we have
\begin{eqnarray}
A^\mu
&=&
\bar{u}(1')k_{2}^{\alpha}G_{\alpha\beta}(p_{1}+Q)
\frac{C_3^V}{m_N}
(g^{\beta\mu}\Qbar-Q^{\beta}\gamma^{\mu})\gamma_5
u(1) 
\nonumber\\
\label{amubis}\\
B^\mu
&=&
\bar{u}(1')k_{2}^{\beta}
\frac{C_3^V}{m_N}\gamma_5
(g^{\alpha\mu}\Qbar-Q^{\alpha}\gamma^{\mu})
G_{\alpha\beta}(p'_{1}-Q)u(1). 
\nonumber\\
\label{bmubis}
\end{eqnarray}
In the last equation
we have permuted the \(\gamma_5\) matrix, which introduces a minus sign that cancels with the negative sign from \(-Q\).
We only need to perform the non-relativistic reduction of the spatial
components (\(\mu = i\)) of the \(\Delta\) current, since we are
computing the transverse response. In the non relativistic limit we 
neglect the time components of $k^\mu$ and $Q^\mu$, i. e.
\begin{equation}
k_2^\mu \simeq (0,\nk_2), \kern 1cm Q^\mu\simeq (0,\nq).
\end{equation} 
Then for the $A^i$ components we have
\begin{eqnarray}
A^i
&\simeq& 
\bar{u}(1')k_{2}^{k}G_{kj}
\frac{C_3^V}{m_N}
(g^{j i}\Qbar-Q^{j}\gamma^{i})\gamma_5
u(1), 
\nonumber\\
&=&
\bar{u}(1')k_{2}^{k}G_{kj}(p_1+Q)\Gamma^{ji}(Q)u(1).
\label{amayusculai}
\end{eqnarray}
Hence at leading order in the non-relativistic limit, only the spatial
components of the \(\Delta\) propagator \( G_{kj} \) and the vertex \(
\Gamma_{ji} \) contribute, while the time components are suppressed.
Analogously, we obtain a similar result for the backward vector
components $B^i$,
\begin{eqnarray}
B^i
&\simeq& 
\bar{u}(1')k_{2}^{k}
\frac{C_3^V}{m_N}
\gamma_5(g^{j i}\Qbar-Q^{j}\gamma^{i})
G_{jk}
u(1), 
\nonumber\\
&=&
\bar{u}(1')  k_{2}^{k}\Gamma^{ji}(-Q) G_{jk}(p'_1-Q) u(1).
\label{bmayusculai}
\end{eqnarray}
Furthermore, the procedure we follow to compute the non-relativistic
reduction of a product of matrix operators is to perform the reduction
on each operator separately. This approach is valid at leading order.

\subsubsection{$\Delta$ propagator}

We begin with the $\Delta$ propagator. In the static limit, with
$p^\mu+Q^\mu \simeq (p^0,0) = (m_N,0)$, and neglecting the lower
components we have
\begin{equation}
\frac{\pbar+m_\Delta}{p^2-m_\Delta^2}
\rightarrow
\frac{p_0+m_\Delta}{p_0^2-m_\Delta^2}
=
\frac{1}{m_N-m_\Delta}.
\end{equation}
Then the $\Delta$ propagator is written as
\begin{eqnarray}
G_{ij} 
&\simeq& -\frac{1}{m_N-m_\Delta}(g_{ij}-\frac13\gamma_i\gamma_j)
\nonumber\\
&\simeq & -\frac{1}{m_N-m_\Delta}(-\delta_{ij}+\frac13\sigma_i\sigma_j)
\nonumber\\
&=& \frac{1}{m_N-m_\Delta}(\frac23\delta_{ij}-i\frac13\epsilon_{ijk}\sigma_k)
\label{delta-prop}
\end{eqnarray}
where we have used the property
\begin{equation}
\sigma_i\sigma_j= i\epsilon_{ijk}\sigma_k
\end{equation}
and $\epsilon_{ijk}$ is the Levi-Civita tensor.

\subsubsection{$\gamma N \Delta$ vertex}

To obtain the non-relativistic reduction of the vertex
\begin{equation}
\Gamma^{ji}(Q)=
\frac{C_3^V}{m_N}
(g^{j i}\Qbar-Q^{j}\gamma^{i})\gamma_5,
\end{equation}
in the low energy limit, we have $Q^\mu\simeq (0,q^i)$. Then 
\begin{eqnarray}
(g^{j i}\Qbar-Q^{j}\gamma^{i})\gamma_5
&\simeq&
\delta_{j i}q^k\gamma^k\gamma_5-q^j\gamma^i\gamma_5
\nonumber\\
&\simeq&
\delta_{j i}q^k\sigma_k-q^j\sigma_i
\nonumber\\
&=&
q^k\sigma_l(\delta_{ij}\delta_{kl}-\delta_{il}\delta_{kj}).
\end{eqnarray}
This expression can be rewritten using the contraction of two
Levi-Civita tensors
\begin{equation}   \label{contraccion}
\epsilon_{ikm}\epsilon_{jlm}=
\delta_{ij}\delta_{kl}-\delta_{il}\delta_{kj}.
\end{equation}
Therefore we have the non relativistic reduction
\begin{equation}
\Gamma^{ji}(Q) \simeq
\frac{C_3^V}{m_N}
(\epsilon_{ikm}q^k)(\epsilon_{jlm}\sigma_l).
\end{equation}

\subsubsection{Forward vector $A^i$}

 From Eq. (\ref{amayusculai}) we have
(we do not write the spinors, just the spin operators):
\begin{eqnarray}
A^i 
&\simeq&
k_2^kG_{kj} 
\frac{C_3^V}{m_N}
(\epsilon_{inm}q^n)(\epsilon_{jlm}\sigma_l)
\nonumber\\
&=&
\frac{C_3^V}{m_N}
\epsilon_{inm}q^n a_m,
\end{eqnarray}
where we have defined the vector
\begin{equation}
a_m \equiv  \epsilon_{jlm} k_2^kG_{kj} \sigma_l.
\end{equation}
(note that $G_{kj}$ and $\sigma_l$ do not commute).
Therefore we can write, in vector form
\begin{equation} \label{Avector}
\nA \simeq \frac{C_3^V}{m_N}(\nq\times\na).
\end{equation}
Hence $\nA$ is purely transverse.

\subsubsection{Backward vector $B^i$}

Similarly,  from Eq. (\ref{bmayusculai}),
\begin{eqnarray}
B^i 
&\simeq&
-k_2^k \frac{C_3^V}{m_N} (\epsilon_{inm}q^n)(\epsilon_{jlm}\sigma_l) G_{jk} 
\nonumber\\
&=&
\frac{C_3^V}{m_N} \epsilon_{inm}q^n b_m,
\end{eqnarray}
where we have defined the vector
\begin{equation}
b_m \equiv
-\epsilon_{jlm}k_2^k  \sigma_l G_{jk}. 
\end{equation}
In vector form we have
\begin{equation} \label{Bvector}
\nB \simeq \frac{C_3^V}{m_N}
(\nq\times\nb).
\end{equation}

\subsubsection{Vectors $a^i$ and $b^i$}

Next, we perform the necessary contractions to derive the explicit
expressions for the vectors \(\na\) and \(\nb\) in the non relativistic limit. 
Using Eq. (\ref{delta-prop}) for the static $\Delta$-propagator,
we have
\begin{eqnarray}
a_m &\simeq& 
\frac{\epsilon_{jlm} k_2^k}{m_N-m_\Delta}
\left(\frac23\delta_{kj}-i\frac13\epsilon_{kjn}\sigma_n\right)\sigma_l
\\
b_m &\simeq& 
\frac{-\epsilon_{jlm} k_2^k}{m_N-m_\Delta}
 \sigma_l \left(\frac23\delta_{kj}-i\frac13\epsilon_{jkn}\sigma_n\right).
\end{eqnarray}
Hence
\begin{eqnarray}
(m_N-m_\Delta) a_m 
&\simeq& 
\frac23  \epsilon_{jlm} k_2^j\sigma_l
\nonumber\\
&&
-\frac{i}3 k_2^k \epsilon_{kjn}\epsilon_{jlm} \sigma_n \sigma_l
\\
(m_N-m_\Delta) b_m 
&\simeq& 
-\frac23  \epsilon_{jlm} k_2^j\sigma_l
\nonumber\\
&&
+\frac{i}3 k_2^k \epsilon_{jkn}\epsilon_{jlm} \sigma_l \sigma_n.
\end{eqnarray}
To compute the contractions in the second summand, we employ 
again the property (\ref{contraccion}) of the Levi-Civita tensor
\begin{equation}
\epsilon_{jnk}\epsilon_{jlm}=\delta_{nl }\delta_{km }-\delta_{nm }\delta_{kl }.
\end{equation}
Then we have
\begin{eqnarray}
k_2^k \epsilon_{kjn}\epsilon_{jlm} \sigma_n \sigma_l
&=&
k_2^k (\delta_{nl }\delta_{km }-\delta_{nm }\delta_{kl })\sigma_n\sigma_l
\nonumber\\
&=& k_2^m\sigma_n\sigma_n-k_2^l\sigma_m\sigma_l
\nonumber\\
&=& 3k_2^m -k_2^l(\delta_{ml}+i\epsilon_{mln}\sigma_n)
\nonumber\\
&=& 2k_2^m -i\epsilon_{mln}k_2^l\sigma_n
\nonumber\\
&=& (2\nk_2 -i\nk_2\times\nsigma)^m,
\end{eqnarray}
and
\begin{eqnarray}
k_2^k \epsilon_{jkn}\epsilon_{jlm} \sigma_l \sigma_n
&=&
k_2^k (\delta_{lk }\delta_{mn }-\delta_{ln }\delta_{mk })\sigma_l\sigma_n
\nonumber\\
&=& k_2^l\sigma_l\sigma_m-k_2^m\sigma_l\sigma_l
\nonumber\\
&=& -3k_2^m + k_2^l(\delta_{lm}+i\epsilon_{lmn}\sigma_n)
\nonumber\\
&=& -2k_2^m -i\epsilon_{mln}k_2^l\sigma_n
\nonumber\\
&=& (-2\nk_2 -i\nk_2\times\nsigma)^m.
\end{eqnarray}
Then we can write in vector form  $\na$ and $\nb$ as
\begin{eqnarray}
(m_N-m_\Delta) \na 
&\simeq& 
\frac23 (\nk_2\times\nsigma) 
-\frac{i}3 (2\nk_2 -i\nk_2\times\nsigma)
\nonumber \\
&=& 
-\frac{2}3 i\nk_2+
\frac13 \nk_2\times\nsigma 
\\
(m_N-m_\Delta) \nb 
&\simeq& 
-\frac23 (\nk_2\times\nsigma) 
+\frac{i}3 (-2\nk_2 -i\nk_2\times\nsigma)
\nonumber\\
&=& 
-\frac{2}3 i\nk_2-\frac13 \nk_2\times\nsigma .
\end{eqnarray}
Using this result in Eqs. (\ref{Avector},\ref{Bvector}) finally we find
\begin{eqnarray} 
\nA &\simeq& \frac{C_3^V}{m_N}
\frac{1}{m_N-m_\Delta}\nq\times
\left[-\frac{2}3 i\nk_2+ \frac13 \nk_2\times\nsigma \right] 
\\
\nB &\simeq& \frac{C_3^V}{m_N}
\frac{1}{m_N-m_\Delta}\nq\times
\left[-\frac{2}3 i\nk_2- \frac13 \nk_2\times\nsigma \right] 
\end{eqnarray}
from where Eqs. (\ref{Anonrel}) and (\ref{Bnonrel}) follow.

\section{Isospin Summations in the 1p1h MEC Matrix Element}

\label{AppendixB}

Here we provide the sums over the isospin index \(t_k\) of the
spectator nucleon appearing in the 1p1h MEC matrix element. 
The isospin dependence of the MEC is of the form
\begin{equation}
\nj= \tau^{(1)}_z \nj_1+ \tau^{(2)}_z \nj_2 + 
i[ \ntau^{(1)} \times \ntau^{(2)}]_z \nj_3,
\end{equation}
where $\ntau^{(1)}$ and $\ntau^{(2)}$ are isospin operators of the
first and second particle, respectively.  We begin by referencing the
Pauli matrices, which also represent the isospin operators required.
\begin{equation}
\tau_1 = \begin{pmatrix}
0 & 1 \\
1 & 0
\end{pmatrix}, \quad
\tau_2 = \begin{pmatrix}
0 & -i \\
i & 0
\end{pmatrix}, \quad
\tau_3 = \begin{pmatrix}
1 & 0 \\
0 & -1
\end{pmatrix}.
\end{equation}

These matrices act on the isospin states of
nucleons, $|t\rangle$,  for protons (\(t =
+\frac{1}{2}\)) and neutrons (\(t = -\frac{1}{2}\)).
We need the basic result
\begin{eqnarray}
\tau_1|p\rangle = |n\rangle,
&&
\tau_1|n\rangle = |p\rangle 
\nonumber \\
i\tau_2|p\rangle = -|n\rangle
&&
i\tau_2|n\rangle = |p\rangle.
\label{tau}
\end{eqnarray}
By expanding the vector product
\begin{equation}
i[\ntau^{(1)} \times \ntau^{(2)}]_z=
i\tau_1^{(1)}\tau_2^{(2)}
-i\tau_2^{(1)}\tau_1^{(2)},
\end{equation}
we obtain
\begin{eqnarray}
i[\ntau^{(1)} \times \ntau^{(2)}]_z  |pp\rangle &=& 0,\nonumber\\
i[\ntau^{(1)} \times \ntau^{(2)}]_z  |nn\rangle &=& 0,\nonumber\\
i[\ntau^{(1)} \times \ntau^{(2)}]_z  |pn\rangle &=& 
2|np\rangle = 4t_p |np\rangle,
\nonumber\\
i[\ntau^{(1)} \times \ntau^{(2)}]_z  |np\rangle &=&
-2|pn\rangle = 4t_n |pn\rangle.
\label{tautau}
\end{eqnarray}
These four equations can be written in unified form as
\begin{equation}
i[\ntau^{(1)} \times \ntau^{(2)}]_z|t_1t_2\rangle=
4t_1 (1-\delta_{t_1t_2})|t_2t_1\rangle.
\end{equation}
From these elementary results, we can compute the 
sums over $t_k$ appearing in the direct and exchange matrix elements of the current. 

\subsubsection{Direct terms.}

For the the direct terms we have:
\begin{equation} \label{iso1}
\sum_{t_k=\pm 1/2} 
\langle t_pt_k|  \tau^{(1)}_z  | t_ht_k\rangle
=
\sum_{t_k} 
     \delta_{t_pt_h}2t_h 
=     \delta_{t_pt_h}4t_h, 
\end{equation}
\begin{equation} \label{iso2}
\sum_{t_k} 
\langle t_pt_k| \tau^{(2)}_z  | t_ht_k\rangle
=
\sum_{t_k} 
 \delta_{t_pt_h}2t_k
= 0,
\end{equation}
\begin{eqnarray}
\sum_{t_k} 
\langle t_pt_k| i[\ntau^{(1)} \times \ntau^{(2)}]_z  | t_ht_k\rangle 
&=& 
\nonumber\\
&& \kern -3cm
= \sum_{t_k} 
\delta_{t_pt_k}\delta_{t_kt_h} 4t_k (1-\delta_{t_ht_k}) = 0.
\label{iso3}
\end{eqnarray}

\subsubsection{Exchange terms.}

For the the exchange matrix elements we have:
\begin{equation} \label{iso4}
\sum_{t_k} 
\langle t_pt_k|  \tau^{(1)}_z  | t_kt_h\rangle
=
\sum_{t_k} 
     \delta_{t_pt_k}
     \delta_{t_kt_h}2t_k 
=     \delta_{t_pt_h}2t_h, 
\end{equation}
\begin{equation} \label{iso5}
\sum_{t_k} 
\langle t_pt_k| \tau^{(2)}_z  | t_kt_h\rangle
=
\sum_{t_k} 
     \delta_{t_pt_k}
     \delta_{t_kt_h}2t_h 
=     \delta_{t_pt_h}2t_h, 
\end{equation}
\begin{eqnarray}
\lefteqn{\sum_{t_k} 
\langle t_pt_k| i[\ntau^{(1)} \times \ntau^{(2)}]_z  | t_kt_h\rangle 
=} 
\nonumber\\
&=& 
\sum_{t_k} 
\langle t_pt_k| 4t_k (1-\delta_{t_kt_h})  | t_ht_k\rangle 
\nonumber\\
&=& \sum_{t_k} 
\delta_{t_pt_h} 4t_k (1-\delta_{t_kt_h}) = 
   - \delta_{t_pt_h}4t_h. 
 \label{iso6}
\end{eqnarray}

\subsubsection{Null $\Delta$ diagrams.}

Next, we will demonstrate that diagrams (e) and (f) corresponding to
the $\Delta$ current are zero after summing over isospin. To achieve
this, we must use the original form of the isospin operators,
Eqs. (\ref{uf},\ref{ub}). The forward current involves the operators
\(U_F(1,2)\) and \(U_F(2,1)\), while the backward current contains the
isospin operators \(U_B(1,2)\) and \(U_B(2,1)\). By carefully
analyzing these operators, we can show that the specific contributions
from diagrams (e) and (f) cancel out, leading to a net zero result for
each.  First, from property (\ref{titj}) we can write the
following products
\begin{eqnarray}
T_1T_3^\dagger = \frac{i}{3}\tau_2
&& 
T_2T_3^\dagger = -\frac{i}{3}\tau_1
\\
T_3T_1^\dagger = -\frac{i}{3}\tau_2
&&T_3T_2^\dagger = \frac{i}{3}\tau_1
\\
T_3T_3^\dagger = \frac{2}{3}.
\end{eqnarray}
From here, using $\tau_1\tau_2=-\tau_2\tau_1=i\tau_3$, we have
\begin{eqnarray}
\sum_i  \tau_i T_iT_3^{\dagger} 
&=& 
\tau_1T_1T_3^\dagger
+\tau_2T_2T_3^\dagger
+\tau_3T_3T_3^\dagger
\nonumber\\
&=&
 \frac{i}{3}\tau_1\tau_2
- \frac{i}{3}\tau_2\tau_1
+ \frac{2}{3}\tau_3
\nonumber\\
&=&
-\frac13\tau_3-\frac13\tau_3+\frac23\tau_3=0.
\end{eqnarray}
In the case of the forward current, the isospin sum of the exchange 
matrix element of the $(1\leftrightarrow 2)$ term is
\begin{eqnarray}
\sum_{t_k}
\langle pk| U_F(2,1) |kh\rangle 
&=&
\sum_{t_k}
\sqrt{\frac32}
\langle pk| \sum_i T_i^{(2)}T_3^{(2)\dagger}\tau_i^{(1)} |kh\rangle
\nonumber\\
&=&\sqrt{\frac32}
\sum_{t_k}\sum_i 
\langle p|\tau_i|k\rangle
\langle k| 
T_iT_3^{\dagger} |h\rangle =
\nonumber\\
&=&
\sqrt{\frac32}\sum_i 
\langle p|\tau_i
T_iT_3^{\dagger} |h\rangle = 0.
\end{eqnarray}
This demonstrates the result for the forward term, that diagram (e) of
Fig. 2 is zero.
 Analogously, the same steps can be applied to show
the result that diagram (f) for the backward term is zero. 
Fist we have
\begin{eqnarray}
\sum_i T_3T_i^\dagger \tau_i &=&
T_3T_1^\dagger\tau_1
+T_3T_2^\dagger\tau_2
T_3T_3^\dagger\tau_3
\nonumber\\
&=&
 -\frac{i}{3}\tau_2\tau_1
+ \frac{i}{3}\tau_1\tau_2
+ \frac{2}{3}\tau_3
\nonumber\\
&=&
-\frac13\tau_3-\frac13\tau_3+\frac23\tau_3=0.
\end{eqnarray}
Then diagram (f) contain the isospin operator $U_B(1,2)$ and
the isospin sum of the exchange matrix element is
\begin{eqnarray}
\sum_{t_k}
\langle pk| U_B(1,2) |kh\rangle 
&=&
\sum_{t_k}
\sqrt{\frac32}
\langle pk| \sum_i T_3^{(1)}T_i^{(1)\dagger}\tau_i^{(2)} |kh\rangle
\nonumber\\
&=&\sqrt{\frac32}
\sum_{t_k}\sum_i 
\langle p| T_3T_i^{\dagger} |k\rangle
\langle k|\tau_i|h\rangle
\nonumber\\
&=&
\sqrt{\frac32}\sum_i 
\langle p|
T_3T_i^{\dagger} \tau_i|h\rangle = 0.
\end{eqnarray}

\begin{widetext}

\section{Spin summations in the 1p1h MEC matrix elements}
\label{AppendixC}

Here we perform the spin summations appearing in the exchange matrix
element, given by
\begin{eqnarray}
\sum_{t_ks_k}\nj_{2b}(p,k,k,h)
&=& \delta_{t_pt_h}2t_h \sum_{s_k}
[\nj_1(p,k,k,h)
+\nj_2(p,k,k,h)-2\nj_3(p,k,k,h)].
\end{eqnarray}

\subsubsection{Seagull current}

In the case of the seagull current only the current $\nj_3$ contribute, given by Eq. (\ref{seagull}). The sum over $t_k,s_k$ is
\begin{eqnarray}
\sum_{t_ks_k}\nj_{s}(p,k,k,h)
&=&
-4t_h\delta_{t_pt_h} 
\sum_{s_k}
\langle s_ps_k|
\frac{f^2}{m_\pi^2}F_1^V
\left(
\frac{\nk_1\cdot\nsigma^{(1)}}{\nk_1^2+m_{\pi}^2}\nsigma^{(2)}
-\frac{\nk_2\cdot\nsigma^{(2)}}{\nk_2^2+m_{\pi}^2}\nsigma^{(1)}
\right)
|s_ks_h\rangle
\nonumber\\
&=&
-4t_h\delta_{t_pt_h} 
\frac{f^2}{m_\pi^2}F_1^V
\sum_{s_k}
\left(
\frac{\nk_1\cdot\nsigma_{pk}}{\nk_1^2+m_{\pi}^2}\nsigma_{kh}
-\frac{\nk_2\cdot\nsigma_{kh}}{\nk_2^2+m_{\pi}^2}\nsigma_{pk}
\right).
\label{seagull2}
\end{eqnarray}
with $\nk_1=\np-\nk$ and $\nk_2=\nk-\nh$.
The separate spin sums are
\begin{equation}
\sum_{s_k}(\nk_1\cdot\nsigma_{pk})\nsigma_{kh}
= (\nk_1+i\nsigma\times\nk_1)_{ph},
\kern 2cm
\sum_{s_k}(\nk_2\cdot\nsigma_{kh})\nsigma_{pk}
= (\nk_2+i\nk_2\times\nsigma)_{ph}.
\end{equation}
We obtain
\begin{eqnarray}
\sum_{t_ks_k}\nj_{s}(p,k,k,h)
&=&
-4t_h\delta_{t_pt_h} 
\frac{f^2}{m_\pi^2}F_1^V
\langle s_p|
\left(
\frac{\nk_1+i\nsigma\times\nk_1}{\nk_1^2+m_{\pi}^2}
-\frac{\nk_2+i\nk_2\times\nsigma}{\nk_2^2+m_{\pi}^2}
\right)
|s_h\rangle.
\end{eqnarray}

\subsubsection{Pionic}

In the case of the pion in flight or pionic current the sum over
spin-isospin reads
\begin{eqnarray}
\sum_{t_ks_k}\nj_{\pi}(p,k,k,h)
&=&
-4t_h\delta_{t_pt_h} 
\sum_{s_k}
\langle s_ps_k|
\frac{f^2}{m_\pi^2}F_1^V
\frac{\nk_1\cdot\nsigma^{(1)}}{\nk_1^2+m_{\pi}^2}
\frac{\nk_2\cdot\nsigma^{(2)}}{\nk_2^2+m_{\pi}^2}(\nk_1-\nk_2)
|s_ks_h\rangle 
\nonumber\\
&=&
-4t_h\delta_{t_pt_h} 
\frac{f^2}{m_\pi^2}F_1^V
\sum_{s_k}
\frac{\nk_1\cdot\nsigma_{pk}}{\nk_1^2+m_{\pi}^2}
\frac{\nk_2\cdot\nsigma_{kh}}{\nk_2^2+m_{\pi}^2}(\nk_1-\nk_2)
\label{pionic2}
\end{eqnarray}
with $k_1=\np-\nk$ and $\nk_2=\nk-\nh$.
The sum over spin index $s_k$ is performed using
\begin{equation}
\sum_{s_k} (\nk_1\cdot\nsigma_{ph})(\nk_2\cdot\nsigma_{kh})=
\nk_1\cdot\nk_2\delta_{s_ps_h}+i(\nk_1\times\nk_2)\cdot\nsigma_{ph} .
\end{equation}

\subsubsection{$\Delta$ current}

From the non-relativistic Eq. (\ref{deltafinal}) we can identify the
three contributions, $\nj_i$, to the $\Delta$ current
\begin{eqnarray}
\sum_{t_ks_k}\nj_{\Delta}(p,k,k,h)
&=&
-i\delta_{t_pt_h}2t_h 
C_\Delta\nq\times 
\sum_{s_k}
\langle s_ps_k|
\frac{\nk_1\cdot\nsigma^{(1)}}{\nk_1^2+m_{\pi}^2}4\nk_1
+\frac{\nk_2\cdot\nsigma^{(2)}}{\nk_2^2+m_{\pi}^2}4\nk_2
\nonumber\\
&&
+2i 
\frac{\nk_2\cdot\nsigma^{(2)}}{\nk_2^2+m_{\pi}^2}
(\nk_2\times\nsigma^{(1)})
-2i
\frac{\nk_1\cdot\nsigma^{(1)}}{\nk_1^2+m_{\pi}^2}
(\nk_1\times\nsigma^{(2)})
|s_ks_h\rangle
\end{eqnarray}
with $\nk_1=\np-\nk$ and $\nk_2=\nk-\nh$.
Writing explicitly the spin indices in the Pauli matrices we have
\begin{eqnarray}
\sum_{t_ks_k}\nj_{\Delta}(p,k,k,h)
&=&
-i\delta_{t_pt_h}2t_h 
C_\Delta\nq\times 
\sum_{s_k}
\left\{
\frac{\nk_1\cdot\nsigma_{pk}}{\nk_1^2+m_{\pi}^2}4\nk_1\delta_{s_ks_h}
+\frac{\nk_2\cdot\nsigma_{kh}}{\nk_2^2+m_{\pi}^2}4\nk_2\delta_{s_ps_k}
\right.
\nonumber\\
&& +2i 
\left.
\frac{\nk_2\cdot\nsigma_{kh}}{\nk_2^2+m_{\pi}^2}
(\nk_2\times\nsigma_{pk})
-2i
\frac{\nk_1\cdot\nsigma_{pk}}{\nk_1^2+m_{\pi}^2}
(\nk_1\times\nsigma_{kh})
\right\}
\end{eqnarray}
with
$\nk_1=\np-\nk$, and $\nk_2=\nk-\nh$.
We need the following spin sums 
\begin{eqnarray}
\sum_{s_k}
(\nsigma_{kh}\cdot\nk_2)(\nk_2\times\nsigma_{pk})
&=&
-i\left[ 
k_2^2\nsigma_{ph}-
 (\nsigma_{ph}\cdot\nk_2)\nk_2
\right]
\nonumber\\
\sum_{s_k}
(\nsigma_{pk}\cdot\nk_1)(\nk_1\times\nsigma_{kh})
&=&
i\left[ 
k_1^2\nsigma_{ph}-
 (\nsigma_{ph}\cdot\nk_1)\nk_1
\right].
\nonumber\\
\end{eqnarray}
The result for the sum over $s_k$ is
\begin{eqnarray}
\sum_{t_ks_k}\nj_{\Delta}(p,k,k,h)
=
-4it_h\delta_{t_pt_h}
C_\Delta\nq\times 
\left\{
\frac{k_1^2\nsigma_{ph}+(\nsigma_{ph}\cdot\nk_1)\nk_1}
     {k_1^2+m_\pi^2}
+\frac{\nk_2^2\nsigma_{ph}+(\nsigma_{ph}\cdot\nk_2)\nk_2}
     {k_2^2+m_\pi^2} .
\right\}
\label{delta_sk}
\end{eqnarray}

\section{Spin summations in the interference responses}
\label{AppendixD}

Here we compute the spin summations in the 1b2b interference response
function.

In this appendix we use the notation $\nk_1=\np-\nk$  and  $\nk_2=\nk-\nh$. 
We use also the identities $\nk_1+\nk_2=\nq$, and $\nk_1=\nq-\nk_2$.

\subsubsection{Magnetization-seagull}

Inserting Eq. (\ref{seagullph}) into Eq. (\ref{mssum}) we have

\begin{eqnarray}
w^T_{ms}
&=&
4t_h\delta_{t_pt_h} 
\frac{G_M^h}{2m_N}
\frac{f^2}{m_\pi^2}F_1^V
\int \frac{d^3k}{(2\pi)^3}
\sum_{s_ps_h}
i(\nq\times\nsigma_{hp})
\cdot 
\left(
\frac{\delta_{s_ps_h}\nk_1+i\nsigma_{ph}\times\nk_1}{\nk_1^2+m_{\pi}^2}
-\frac{\delta_{s_ps_h}\nk_2+i\nk_2\times\nsigma_{ph}}{\nk_2^2+m_{\pi}^2}
\right).
\end{eqnarray}
The sums involved inside the integral are of the kind:
\begin{eqnarray}
\sum_{s_ps_h}
i(\nq\times\nsigma_{hp})\cdot
\delta_{s_ps_h}\nk
&=& 
\sum_{s_h} i(\nq\times\nsigma_{hh})\cdot\nk =0
\\
\sum_{s_ps_h}
i(\nq\times\nsigma_{hp})\cdot
(i\nsigma_{ph}\times\nk)
&=& 
-\sum_{s_ps_h}
(\nq\cdot\nsigma_{ph} \nsigma_{hp}\cdot \nk
-\nq\cdot\nk \nsigma_{hp}\cdot\nsigma_{ph})
=
-\mbox{Tr}(q^i\sigma_i\sigma_j k^j-\nq\cdot\nk \sigma_i\sigma_i)
\nonumber\\
&=&
-\mbox{Tr}(q^i\delta_{ij} k^j-3\nq\cdot\nk)=4\nq\cdot\nk .
\label{qssk}
\end{eqnarray}
Therefore
\begin{eqnarray}
w^T_{ms}
&=&
4t_h\delta_{t_pt_h} 
\frac{G_M^h}{2m_N}
\frac{f^2}{m_\pi^2}F_1^V
\int \frac{d^3k}{(2\pi)^3}
\left(
\frac{4\nq\cdot\nk_1}{\nk_1^2+m_{\pi}^2}
+\frac{4\nq\cdot\nk_2}{\nk_2^2+m_{\pi}^2}
\right).
\end{eqnarray}

\subsubsection{Convection-seagull}

Inserting Eq. (\ref{seagullph}) into Eq. (\ref{cssum}) we have

\begin{eqnarray}
w^T_{cs} 
& = & 
4t_h\delta_{t_pt_h} 
\frac{G_E^h}{m_N} 
\frac{f^2}{m_\pi^2}F_1^V
\int \frac{d^3k}{(2\pi)^3}
\sum_{s_ps_h}
\delta_{s_hs_p} \nh_T \cdot 
\left(
\frac{\delta_{s_ps_h}\nk_1+i\nsigma_{ph}\times\nk_1}{\nk_1^2+m_{\pi}^2}
-\frac{\delta_{s_ps_h}\nk_2+i\nk_2\times\nsigma_{ph}}{\nk_2^2+m_{\pi}^2}
\right).
\nonumber\\
& = & 
4t_h\delta_{t_pt_h} 
\frac{G_E^h}{m_N} 
\frac{f^2}{m_\pi^2}F_1^V
\int \frac{d^3k}{(2\pi)^3}
\left(
\frac{\nh_T \cdot \nk_1}{\nk_1^2+m_{\pi}^2}
-\frac{\nh_T \cdot \nk_2}{\nk_2^2+m_{\pi}^2}
\right).
\end{eqnarray}

\subsubsection{Magnetization-pionic}

Inserting Eq. (\ref{pionicph}) into Eq. (\ref{mpisum}) we have

\begin{eqnarray}
w^T_{m\pi}&  = & 
4t_h\delta_{t_pt_h} 
\frac{G_M^h}{2m_N}
\frac{f^2}{m_\pi^2}F_1^V
\int \frac{d^3k}{(2\pi)^3}
\sum_{s_ps_h}
i(\nq\times\nsigma_{hp})
\cdot (\nk_1-\nk_2)
\frac{\delta_{s_ps_h}\nk_1\cdot\nk_2
+i(\nk_1\times\nk_2)\cdot\nsigma_{ph}}
{(\nk_1^2+m_{\pi}^2)(\nk_1^2+m_{\pi}^2)}.
\end{eqnarray}
The sum over spin inside the integral is
\begin{eqnarray}
\sum_{s_ps_h}
i(\nq\times\nsigma_{hp})
\cdot (\nk_1-\nk_2)
\left[ \delta_{s_ps_h}\nk_1\cdot\nk_2
       +i(\nk_1\times\nk_2)\cdot\nsigma_{ph}
\right]
=
-\sum_{s_ps_h}
(\nq\times\nsigma_{hp})
\cdot (\nk_1-\nk_2)
(\nk_1\times\nk_2)\cdot\nsigma_{ph}
\nonumber\\
=
-\sum_{s_ps_h}
[(\nk_1-\nk_2)\times\nq]\cdot\nsigma_{hp}
(\nk_1\times\nk_2)\cdot\nsigma_{ph}
=
-2
[(\nk_1-\nk_2)\times\nq]\cdot
(\nk_1\times\nk_2)
\nonumber\\
=-4(\nq\times\nk_2)^2
 \label{trazampi}
\end{eqnarray}
where we have used that 
\begin{eqnarray}
\sum_{s_ps_h}(\na\cdot\nsigma_{hp})
(\nb\cdot\nsigma_{ph})=
{\mbox Tr}(a^i\sigma_i\sigma_jb^j)=2\na\cdot\nb
\label{asigbsig}\\
\nk_1=\nq-\nk_2
\\
(\nk_1-\nk_2)\times\nq=
(\nq-2\nk_2)\times\nq=2\nq\times\nk_2
\\
\nk_1\times\nk_2=
(\nq-\nk_2)\times\nk_2=
\nq\times\nk_2 .
\end{eqnarray}
With the result of Eq. (\ref{trazampi}), we  obtain
\begin{eqnarray}
w^T_{m\pi}&  = & 
4t_h\delta_{t_pt_h} 
\frac{G_M^h}{2m_N}
\frac{f^2}{m_\pi^2}F_1^V
\int \frac{d^3k}{(2\pi)^3}
\frac{-4(\nq\times\nk_2)^2}
{(\nk_1^2+m_{\pi}^2)(\nk_1^2+m_{\pi}^2)}.
\end{eqnarray}

\subsubsection{Convection-pionic}

Inserting Eq. (\ref{pionicph}) into Eq. (\ref{cpisum}) we have

\begin{eqnarray}
w^T_{c\pi} & = & 
4t_h\delta_{t_pt_h} 
\frac{G_E^h}{m_N} 
\frac{f^2}{m_\pi^2}F_1^V
\int \frac{d^3k}{(2\pi)^3}
\sum_{s_ps_h}
\delta_{s_hs_p} \nh_T \cdot (\nk_1-\nk_2)
\frac{\delta_{s_ps_h}\nk_1\cdot\nk_2
+i(\nk_1\times\nk_2)\cdot\nsigma_{ph}}
{(\nk_1^2+m_{\pi}^2)(\nk_1^2+m_{\pi}^2)}.
\end{eqnarray}
Sum over spin inside the integral:
\begin{eqnarray}
\sum_{s_ps_h}
\delta_{s_hs_p} \nh_T \cdot (\nk_1-\nk_2)
[\delta_{s_ps_h}\nk_1\cdot\nk_2
+i(\nk_1\times\nk_2)\cdot\nsigma_{ph}]
=
\sum_{s_h}
\nh_T \cdot (\nk_1-\nk_2)(\nk_1\cdot\nk_2)
\nonumber\\
=
2 \nh_T \cdot (\nq-2\nk_2)[(\nq-\nk_2)\cdot\nk_2]
=
-4 (\nh_T \cdot\nk_2)(\nq\cdot\nk_2-\nk_2^2).
\end{eqnarray}
Then we obtain
\begin{eqnarray}
w^T_{c\pi} & = & 
4t_h\delta_{t_pt_h} 
\frac{G_E^h}{m_N} 
\frac{f^2}{m_\pi^2}F_1^V
\int \frac{d^3k}{(2\pi)^3}
\frac{
-4 (\nh_T \cdot\nk_2)(\nq\cdot\nk_2-\nk_2^2)
}
{(\nk_1^2+m_{\pi}^2)(\nk_1^2+m_{\pi}^2)}.
\end{eqnarray}

\subsubsection{Magnetization-$\Delta$}

Inserting Eq. (\ref{deltaph}) into Eq. (\ref{mdsum}) we have
\begin{equation}
w^T_{m\Delta}  =  
4t_h\delta_{t_pt_h}
\frac{G_M^h}{2m_N}
C_\Delta
\int \frac{d^3k}{(2\pi)^3}
\sum_{s_ps_h}
i(\nq\times\nsigma_{hp})
\cdot 
\left[
i\nq\times 
\left(
\frac{\nk_1^2\nsigma_{ph}+(\nsigma_{ph}\cdot\nk_1)\nk_1}
     {\nk_1^2+m_\pi^2}
+\frac{\nk_2^2\nsigma_{ph}+(\nsigma_{ph}\cdot\nk_2)\nk_2}
     {\nk_2^2+m_\pi^2}
\right)
\right].
\end{equation}
We need the following spin sums. The first one is similar to Eq. (\ref{qssk})
\begin{eqnarray}
\sum_{s_ps_h}
(\nq\times\nsigma_{hp})
\cdot 
(\nq\times\nsigma_{ph}) 
=4q^2.
\end{eqnarray}
The second sum required is
\begin{eqnarray}
\sum_{s_ps_h}(\nq\times\nsigma_{hp})
\cdot (\nq\times \nk_1)
(\nsigma_{ph}\cdot\nk_1)
&=&
\sum_{s_ps_h}
[q^2 \nsigma_{hp}\cdot \nk_1-(\nq\cdot\nk_1)(\nq\cdot\nsigma_{hp})]
(\nsigma_{ph}\cdot\nk_1)
\nonumber\\
&&
\kern -2cm
=q^2 \sum_{s_ps_h}(\nsigma_{hp}\cdot \nk_1) (\nsigma_{ph}\cdot\nk_1)
-(\nq\cdot\nk_1)\sum_{s_ps_h}(\nq\cdot\nsigma_{hp})(\nsigma_{ph}\cdot\nk_1)
\nonumber\\ 
&=&
2q^2 k_1^2 -2(\nq\cdot\nk_1)^2
\end{eqnarray}
where we have used twice Eq. (\ref{asigbsig}).
Using these results the $m\Delta$ response function is
\begin{equation}
w^T_{m\Delta}  =  
-4t_h\delta_{t_pt_h}
\frac{G_M^h}{2m_N}
C_\Delta
\int \frac{d^3k}{(2\pi)^3}
\left(
\frac{6q^2k_1^2-2(\nq\cdot\nk_1)^2}
     {\nk_1^2+m_\pi^2}
+\frac{6q^2k_2^2-2(\nq\cdot\nk_2)^2}
     {\nk_2^2+m_\pi^2}
\right).
\end{equation}

\end{widetext}

\section{Spectral function and hadronic tensor}
\label{AppendixE}

The spectral function is obtained by assuming plane waves for the
final nucleon. This assumption leads to a factorization approximation
for the response function within the impulse approximation, where the
current is considered to be one-body only. In this framework, the
response function can be factored into a product of the current matrix
element and the spectral function, which describes the distribution of
hole states in the nucleus. 

We assume that the initial nuclear state is a spin-zero nucleus at rest
with energy $E_i=M_A$, and wave function:
\begin{equation}
| i \rangle = | \Phi_0^{(A)} \rangle.
\end{equation}
The final state correspond to a plane wave particle and a residual $A-1$ nucleus  
\begin{equation}
| f \rangle = | \Phi_{\alpha}^{(A-1)}, \np,s \rangle= a^\dagger_{\np,s} 
| \Phi_{\alpha}^{(A-1)} \rangle.
\end{equation}
The label $\alpha$ denotes the quantum numbers of the daughter nucleus
in an excited state with excitation energy
$\epsilon_\alpha^{(A-1)}$. Then the final energy, neglecting the
recoil energy is,
\begin{equation}
E_f=m_N+ T_p+ M_{A-1}+ \epsilon_\alpha^{(A-1)}
\end{equation}
where $T_p=p^2/2m_N$. Then the diagonal component of the hadronic tensor is,
\begin{eqnarray}
W^{\mu\mu} &=& 
\sum_{\alpha\np s} |\langle \Phi_{\alpha}^{(A-1)}, \np,s|J^{\mu}(\nq) | \Phi_0^{(A)} \rangle|^2
\delta(E_i+\omega-E_f). \nonumber \\
\end{eqnarray}
Assuming that the current is a one-body operator and ignoring the
final nucleon spin for simplicity we have,
\begin{eqnarray}
W^{\mu\mu}
&=& \nonumber \\
&&
\kern -1cm 
\sum_{\alpha\np} |\langle \Phi_{\alpha}^{(A-1)}| a_{\np} \int d^3k
J^{\mu}(\nq+\nk,\nk) a^\dagger_{\nq+\nk}a_{\nk}| \Phi_0^{(A)}
\rangle|^2 \nonumber \\
 && 
\kern -1cm
\delta(M_A + \omega -m_N- T_p- M_{A-1}-
\epsilon_\alpha^{(A-1)}).
\end{eqnarray}
Using the commutation properties of the creation and annihilation
operators,
\begin{equation}
a_{\np} a^\dagger_{\nq+\nk}= \delta(\np-\nq-\nk)-a^\dagger_{\nq+\nk}a_{\np},
\end{equation}
and assuming that the final particle momentum is large enough to neglect
high-momentum components in the initial wave function (as the dominant
contribution comes from momenta below the Fermi momentum),
$a_{\np}| \Phi_0^{(A)}\rangle \simeq 0$, then the hadronic tensor is
\begin{eqnarray}
W^{\mu\mu}
&=& \nonumber \\
&&
\kern -1cm 
\sum_{\alpha}\int d^3p |\langle \Phi_{\alpha}^{(A-1)}|
J^{\mu}(\np,\np-\nq) a_{\np-\nq}| \Phi_0^{(A)} \rangle|^2 \nonumber \\
 && 
\kern -1cm
\delta(M_A + \omega -m_N- T_p- M_{A-1}-
\epsilon_\alpha^{(A-1)}).
\end{eqnarray}
Introducing the separation energy $S=M_{A-1}+m_N-M_A>0$ and the missing energy $E_m=\omega-T_p$ then
\begin{equation}
W^{\mu\mu}=
\int d^3p |J^{\mu}(\np,\np-\nq) |^2 S(\np-\nq,E_m)\nonumber \\
\end{equation}
where the one-hole spectral function is defined as,
\begin{equation}
S(\nh,E) = \sum_{\alpha,s} |\langle \Phi_{\alpha}^{(A-1)}|
a_{\nh,s}| \Phi_0^{(A)} \rangle|^2 
\delta(E-S-\epsilon_\alpha^{(A-1)}). \nonumber \\
\end{equation}

\section{Relativistic Mean Field}
\label{AppendixG}

In this appendix we compare the expression of the relativistic
$\Delta$ current with the one used in the RMF model of 
Refs. \cite{Fra23,Fra25}.

In the relativistic mean field for finite nucleus the wave function of
the hole in momentum space is $\langle \np |h\rangle = \Psi_h (\np) $,
while $\langle \np |\np_N\rangle = \Psi_{\np_N} (\np)$ is the one for
the particle, where $\np_N$ is the asymptotic momentum of the final nucleon
The two-body current matrix element is the sum of direct, $D^\mu$,
plus exchange, $E^\mu$ parts  
\begin{eqnarray}
\left\langle ph^{-1} \right|\hat{J}_{2b}^{\mu} |\left. C \right\rangle 
&=&
\sum_{k}\left[
\left\langle pk \right|\hat{J}_{2b}^{\mu} |\left. hk \right\rangle 
- \left\langle pk \right|\hat{J}_{2b}^{\mu} |\left. kh \right\rangle
\right]
\nonumber\\
&\equiv&
D^\mu+E^\mu,
\end{eqnarray}
where $|k\rangle $ is a bound state wave function $\Psi_K(\nr)$. 
We consider the exchange term.
\begin{widetext}
\begin{eqnarray}
  E^\mu &=&
  -\sum_{k}
\left\langle pk \right|\hat{J}_{2b}^{\mu} |\left. kh \right\rangle
=
  -\sum_{k}
\int d^3p'_1 \int d^3p'_2
\int d^3p_1 \int d^3p_2
\langle \np_N k | \np'_1\np'_2\rangle
\langle \np'_1\np'_2| J^{\mu}(\nq) |\np_1\np_2\rangle
\langle \np_1\np_2|
k h \rangle
\nonumber\\
&=&
  -\sum_{k}
\int d^3p'_1 \int d^3p'_2
\int d^3p_1 \int d^3p_2
\overline{\Psi}_{\np_N}(\np'_1)
\overline{\Psi}_{k}(\np'_2)
\frac{\delta(\np_1+\np_2+\nq-\np'_1-\np'_2)}{(2\pi)^3}
 J^{\mu}(\np'_1,\np'_2,\np_1,\np_2)
\Psi_k(\np_1)\Psi_h(\np_2)
\nonumber\\
&=&
 -\sum_{k}
\frac{1}{(2\pi)^3}
\int d^3p'_2
\int d^3p_1 \int d^3p_2
\overline{\Psi}_{\np_N}(\np'_1)
\overline{\Psi}_{k}(\np'_2)
 J^{\mu}(\np'_1,\np'_2,\np_1,\np_2)
\Psi_k(\np_1)\Psi_h(\np_2)
\end{eqnarray}
with $\np'_1=\np_1+\np_2+\nq-\np'_2$, and we have used Eq. (\ref{TBmatrix})
with the full-space normalization $V\rightarrow (2\pi)^3$.
In Ref. \cite{Fra23}, the individual matrix elements for all direct
and exchange MEC diagrams are provided using a different notation. To
facilitate comparison, we consider the first term of the forward current, as given in
Eq. (\ref{delta1}):
\[
  j^{\mu}_{\Delta F1}(p'_1,p'_2,p_1,p_2)=
U_F(1,2)
\frac{ff^{*}}{m_{\pi}^{2}}
V(2',2)
F_{\pi N \Delta}(k_{2}^{2})
\bar{u}_{s'_1}(p'_{1})k_{2}^{\alpha}G_{\alpha\beta}(p_{1}+Q)
\Gamma^{\beta\mu}(Q)u_{s_1}(p_{1}). 
\]
The Exchange matrix element of this current is
\begin{eqnarray}
E^\mu_{\Delta F1} &=&
 -\sum_{k} \frac{1}{(2\pi)^3}\int d^3p'_2 \int d^3p_1 \int d^3p_2
\overline{\Psi}_{\np_N}(\np'_1)
\overline{\Psi}_{k}(\np'_2)
 J^{\mu}(\np'_1,\np'_2,\np_1,\np_2)
\Psi_k(\np_1)\Psi_h(\np_2)
\nonumber\\
&=&
 -\sum_{k} \frac{1}{(2\pi)^3}\int d^3p'_2 \int d^3p_1 \int d^3p_2
\langle p_N k |U_F(1,2) |kh \rangle
\frac{ff^{*}}{m_{\pi}^{2}}
 F_{\pi NN}(k_{2}^{2})
\frac{\bar{\Psi}_{k}(p'_{2})\gamma^{5}\kbar_{2}\Psi_{h}(p_{2})}{k_2^2-m_{\pi}^2}
\nonumber\\
&&
F_{\pi N \Delta}(k_{2}^{2})
\bar{\Psi}_{\np_N}(p'_{1})k_{2}^{\alpha}G_{\alpha\beta}(p_{1}+Q)
\Gamma^{\beta\mu}(Q)\Psi_{k}(p_{1}) 
\nonumber\\
&=&
 -\frac{1}{(2\pi)^3}\int d^3p'_2 \int d^3p_1 \int d^3p_2
\langle p_N k |U_F(1,2) |kh \rangle
\frac{ff^{*}}{m_{\pi}^{2}}
 F_{\pi NN}(k_{2}^{2})F_{\pi N \Delta}(k_{2}^{2})
\nonumber\\
&&
\bar{\Psi}_{\np_N}(p'_{1})k_{2}^{\alpha}G_{\alpha\beta}(p_{1}+Q)
\Gamma^{\beta\mu}(Q)
\frac{\sum_k\Psi_{k}(p_{1}) \bar{\Psi}_{k}(p'_{2})
  }{k_2^2-m_{\pi}^2}
\gamma^{5}\kbar_{2}\Psi_{h}(p_{2})
\nonumber\\
&=&
 \frac{1}{(2\pi)^3}\int d^3p'_2 \int d^3p_1 \int d^3p_2
 \bar{\Psi}_{\np_N}(p'_{1})
 \Gamma^\mu_{\Delta F1}
\Psi_{h}(p_{2}),
\end{eqnarray}
where the $\Gamma^\mu_{\Delta F1}$ operator inside the integral is
defined in a similar manner to that used in Ref. \cite{Fra23}:
\begin{equation}
   \Gamma^\mu_{\Delta F1}=
-\langle p_N k |U_F(1,2) |kh \rangle
\frac{ff^{*}}{m_{\pi}^{2}}
F_{\pi NN}(k_{2}^{2})F_{\pi N \Delta}(k_{2}^{2})
k_{2}^{\alpha}G_{\alpha\beta}(p_{1}+Q)
\Gamma^{\beta\mu}(Q)
\frac{\Lambda_{N'}(p_1,p'_2)
  }{k_2^2-m_{\pi}^2}
\gamma^{5}\kbar_{2}.
\label{casale}
\end{equation}
The function $\Lambda_{N'}(p_1,p'_2) \equiv \sum_k\Psi_{k}(p_{1})
\bar{\Psi}_{k}(p'_{2})$, is defined as in ref. \cite{Fra23} as well (the
sum over k does not include isospin).
Using Eqs. (\ref{tau},\ref{tautau})
the different isospin matrix elements 
in  the $\Delta$F1  exchange term are
$
 I_{\Delta F1}\equiv \langle p_N k | U_F(1,2) |k h \rangle 
 $
 are the following.

 Case $h,p_N= $ proton:
\begin{eqnarray}
k=\mbox{proton} \Rightarrow  && I_{\Delta F1}=
\langle pp |
      \frac{1}{\sqrt{6}}
      \left(2\tau_{z}^{(2)} -i[\ntau^{(1)}\times\ntau^{(2)}]_{z}\right)
| p p  \rangle  
=
\langle pp | \frac{1}{\sqrt{6}} 2| p p  \rangle  
=\frac{2}{\sqrt{6}} = \sqrt{\frac23}
\nonumber\\
k=\mbox{neutron} \Rightarrow  && I_{\Delta F1}=
\langle pn |
      \frac{
      2\tau_{z}^{(2)} -i[\ntau^{(1)}\times\ntau^{(2)}]_{z}}{\sqrt{6}}
| n p  \rangle  
=
\langle pn | \frac{1}{\sqrt{6}}( 2| np \rangle + 2| p n  \rangle)  
 = \sqrt{\frac23}
\nonumber
\end{eqnarray}
Case $h,p_N=$ neutron:
\begin{eqnarray}
k=\mbox{proton} \Rightarrow && I_{\Delta F1}=
\langle np |
      \frac{
      2\tau_{z}^{(2)} -i[\ntau^{(1)}\times\ntau^{(2)}]_{z}}{\sqrt{6}}
| pn  \rangle  
=
\langle np | \frac{1}{\sqrt{6}}( -2| pn \rangle - 2| np  \rangle)  
= -\sqrt{\frac23}
\nonumber\\
k=\mbox{neutron}\Rightarrow && I_{\Delta F1}=
\langle nn |
      \frac{1}{\sqrt{6}}
      \left(2\tau_{z}^{(2)} -i[\ntau^{(1)}\times\ntau^{(2)}]_{z}\right)
| nn  \rangle  
=
\langle nn |\frac{1}{\sqrt{6}} (-2)| nn  \rangle  
= -\sqrt{\frac23}
\nonumber
\end{eqnarray}
Equation (\ref{casale}) must be compared with Eq. (20) of
Ref. \cite{Fra23}:
\begin{equation}
   \Gamma^\mu_{\Delta (b)}=
-I F_{\pi NN}F_{\pi \Delta N}
\frac{f}{m_{\pi}}
\Gamma^\alpha_{\Delta\pi N}
S_{\Delta,\alpha\beta}
\Gamma^{\beta\mu}_{\gamma\Delta N}
\frac{\Lambda_{N'}(p_h,p_p)
  }{k_\pi^2-m_{\pi}^2}
\kbar_{\pi}\gamma^{5},
\kern 1cm
P_{\Delta,b}=Q+P_h,
\label{franco}
\end{equation}
By comparing (\ref{casale}) and (\ref{franco}), we can identify the
correspondence between the different factors appearing in each of
them:
\begin{equation}
 k_\pi=k_2, \kern 1cm P_h=p_1, \kern 1cm
  \Gamma^\alpha_{\Delta\pi N}=\sqrt{2}\frac{f^*}{m_\pi}k^\alpha_2,
  \kern 1cm
  S_{\Delta,\alpha\beta}=G_{\alpha\beta}(p_{1}+Q),
  \kern 1cm
  \Gamma^{\beta\mu}_{\gamma\Delta N}=\Gamma^{\beta\mu}(Q)
\nonumber
\end{equation}
Introducing  the isospin values $ I_{\Delta(b)}=\pm \frac{1}{\sqrt{3}}$ from table I of ref. \cite{Fra23} and $
I_{\Delta F1}=\pm \sqrt{\frac{2}{3}}$, we see that Eqs. (\ref{franco})
and (\ref{casale}) only differ in the last factors
$\kbar_{\pi}\gamma^{5}$ and $\gamma^{5}\kbar_{2}$, respectively, which are opposite
in sign. Therefore
\begin{equation}
\Gamma^\mu_{\Delta (b)}= -  \Gamma^\mu_{\Delta F1}.
\end{equation}
The proof that the remaining terms of the $\Delta$ current in
Ref. \cite{Fra23} have the opposite sign compared to ours can be
carried out in a similar manner.  It can also be seen that the seagull
and pion-in-flight currents in Ref. \cite{Fra23}, by contrast, agree
with those used in the present work.

\end{widetext}


\end{document}